\documentclass[12pt]{article}

\usepackage{amssymb}
\usepackage{amsmath}
\usepackage{amscd}
\usepackage{latexsym}
\usepackage{graphicx}

\newcommand{\be}{\begin{equation}}
\newcommand{\ee}{\end{equation}}

\newcommand{\dlt}{\delta}

\newcommand{\bt}{\beta}

\newcommand{\ep}{\varepsilon}

\newcommand{\ra}{\rightarrow}

\newcommand{\gm}{\gamma}
\newcommand{\om}{\omega}

\newcommand{\lbd}{\lambda}

\newcommand{\cH}{{\cal H}}

\newcommand{\cL}{{\cal L}}
\newcommand{\rgl}{\rangle}
\newcommand{\lgl}{\langle}

\begin{document}

\begin{center}

{\Large{\bf Role of information in decision making of social agents} \\ [5mm]

V.I. Yukalov$^{a,b*}$ and D. Sornette$^{a,c}$} \\ [3mm]

{\it
$^a$Department of Management, Technology and Economics, \\
ETH Z\"urich, Swiss Federal Institute of Technology, \\
Z\"urich CH-8092, Switzerland \\ 
e-mail: syukalov@ethz.ch \\ [3mm]

$^b$Bogolubov Laboratory of Theoretical Physics, \\
Joint Institute for Nuclear Research, Dubna 141980, Russia \\
e-mail: yukalov@theor.jinr.ru \\ [3mm]

$^c$Swiss Finance Institute, c/o University of Geneva, \\
40 blvd. Du Pont d'Arve, CH 1211 Geneva 4, Switzerland \\
e-mail: dsornette@ethz.ch}

\end{center}

\vskip 1cm

\begin{abstract}

The influence of additional information on the decision making of agents, who are 
interacting members of a society, is analyzed within the mathematical 
framework based on the use of quantum probabilities. The introduction of 
social interactions, which influence the decisions of individual agents, 
leads to a generalization of the quantum decision theory developed earlier 
by the authors for separate individuals. The generalized approach is free 
of the standard paradoxes of classical decision theory. This approach also 
explains the error-attenuation effects observed for the paradoxes occurring 
when decision makers, who are members of a society, consult with each 
other, increasing in this way the available mutual information. A precise 
correspondence between quantum decision theory and classical utility theory 
is formulated via the introduction of an intermediate probabilistic version 
of utility theory of a novel form, which obeys the requirement that 
zero-utility prospects should have zero probability weights.
\end{abstract}

\vskip 1cm
   
{\parindent=0pt


\vskip 5mm


\vskip 2cm

$^*$Corresponding author: V.I. Yukalov\\
{\it E-mail address}: yukalov@theor.jinr.ru}
 
\newpage

\section{Paper goal and guide}

The main goal of the paper is to develop a mathematical model describing 
the role of additional information received by decision makers 
characterized by two features. First, the decision makers take decisions not
according to the normative utility theory, but their decisions are influenced 
by behavioral biases. Second, these decision makers are members of a society,
where they interact with each other by exchanging information that influences 
their decisions. 

To treat the first problem of including behavioral biases into decision making,
we employ the approach we recently developed, based on the use of quantum 
probabilities (Yukalov and Sornette, 2008, 2009a,b,c, 2010a,b, 2011).
In this approach, we employ the techniques similar to those used in quantum 
theory, justifying the name ``Quantum Decision Theory'' (QDT). This, however, 
does not require that decision makers be quantum objects, but the used 
techniques just play the role of a convenient mathematical tool for characterizing behavioral biases. It is well known that classical decision making, based on 
the notion of utility theory, is plagued by numerous paradoxes caused by 
behavioral biases. We have shown in our previous publications that in our 
approach, all these paradoxes are cured, since QDT takes behavioral biases into 
account.

In our previous papers, QDT has been developed for single decision makers 
isolated from others and taking decisions on the basis of their present available information. But when decision makers are the agents of a society,
they interact with each other by exchanging information. The principal 
problem, we consider in the present paper, is the generalization of the QDT
approach, taking account of behavioral biases, to a society of interacting agents. 

All the technical, as well as the related psychological points are 
carefully explained in the following sections. However, because the mathematical 
methods we use may be not customary for scholars in social sciences, in 
order that they would not be lost in technical details, it can be useful to 
give a brief guide of the particular problems treated in each of the sections.

In Section 2, we explain why it has been necessary to develop a novel 
approach in order to avoid numerous behavioral paradoxes of classical decision 
making, based on utility theory. We recall that the mathematical approach, 
we advocate, contains no paradoxes when considering isolated decision 
makers. And we stress that there are other interesting phenomena related
to behavioral biases, when decision makers are the members of a society and
interact with each other. The most known among such phenomena is the 
{\it bias attenuation effect}, when the exchange of information between the 
society members reduces the role of behavioral biases.

In Section 3, we generalize QDT to the case of social decision makers, who
are members of a society and who interact with each other. Developing any novel 
theory, it is crucially important to understand under what conditions it can 
be reduced to the known theories. We show that classical decision theory 
is a limiting case of QDT, corresponding to the neglect of behavioral biases
in a precise sense that can be expressed rigorously in mathematical language.
It is also very important that the developed theory would not be just a 
mathematical exercise, but that it could be applied to real-life situations 
and would explain them. We illustrate this by explaining how QDT is used for 
removing the disjunction effect and conjunction error, making predictions 
that are in agreement with empirical observations.  

In Section 4, we analyze how the information received by an agent from other
members of the society influences his/her decisions by changing the role 
of behavioral biases. We explain the conjunction fallacy attenuation observed 
in real-life experiments.      
  
In Section 5, we summarize the novel results of the present paper and stress
the basic problems it has solved. Some possible future experimental studies 
are discussed. The main questions and related answers are explicitly listed,
providing to the reader a clear understanding of the paper content.

\section{Importance of behavioral biases in decision making}

Decision theory underlies essentially all the social sciences, including
economics, finance, political sciences, psychology, and so on. It is also 
employed in studying the evolution of various social systems, where the 
evolution equations that describe population dynamics are constructed so 
as to provide the maximum of utility, or fitness, for the species of 
the considered social system. Decision theory is also an important part
of information technology, including quantum information processing.
 
Decision makers are usually members of a society and, hence, are influenced 
by other members of the society through mutual exchange of information. This 
is why information processing and decision making are intimately connected 
with each other. Recent trends in the interconnection of these subjects have 
been emphasized by Shi (2009). 

The predominant theory, describing individual behavior under risk and uncertainty 
is nowadays the expected utility theory of preferences over uncertain prospects.
This theory, first introduced by Bernoulli (1738) in his investigation of the 
St. Petersburg paradox, was axiomatized by von Neumann and Morgenstern (1953), 
and integrated with the theory of subjective probability by Savage (1954). The
theory was shown to possess great analytical power by Arrow (1971) and Pratt 
(1964) in their work on risk aversion and by Rothschild and Stiglitz (1970, 1971) 
in their work on comparative risk. Friedman and Savage (1948) and Markowitz (1952)
demonstrated its tremendous flexibility in representing decision makers attitudes 
toward risk. It is fair to state that expected utility theory has provided 
a solid foundation for the theory of games, the theory of investment and 
capital markets, the theory of search, and for other branches of economics, 
finance, and management (Lindgren, 1971; White, 1976; Hastings and Mello, 1978;
Rivett, 1980; Buchanan, 1982; Berger, 1985; Marshall and Oliver, 1995; Bather,
2000; French and Insua, 2000; Raiffa and Schlaifer, 2000; Weirich, 2001; 
Gollier, 2001).

However, a number of economists and psychologists have uncovered a growing 
body of evidence showing that individuals do not always conform to prescriptions 
of expected utility theory. Moreover, human beings very often depart from the
theory in predictable and systematic way. Actually, the possibility that problems
could arise has already been discussed by Bernoulli (1738) himself. Then, 
many researchers, starting with the works by Allais (1953), Edwards (1955, 1962),
and Ellsberg (1961), and continuing through the present, have experimentally
confirmed pronounced and systematic deviations from the predictions of expected
utility theory, leading to the appearance of many paradoxes. Neuroscience research
suggests that the choice process used by human beings is systematically biased 
and suboptimal (Fehr and Rangel, 2011). Among the known paradoxes of classical
utility making, we can list the Bernoulli St. Petersburg paradox (Bernoulli, 1738),
the Allais paradox (Allais, 1953), the independence paradox (Allais, 1953), the
Ellsberg paradox (Ellsberg, 1961), the Kahneman-Tversky paradox (Kahneman 
and Tversky, 1979), the Rabin paradox (Rabin, 2000), the Ariely paradox (Ariely,
2008), the disjunction effect (Tversky and Shafir, 1992), the conjunction fallacy
(Tversky and Kahneman, 1983; Shafir et al., 1990), the isolation effects 
(McCaffery and Baron, 2006), the combined paradoxes (Yukalov and Sornette, 2009b,
2010b, 2011), the planning paradox (Kydland and Prescott, 1977), and dynamic
inconsistency (Strotz, 1955; Frederick et al., 2002). A large literature on this
topic can be found in the recent reviews (Camerer et al., 2003; Machina, 2008).

All paradoxes, which have been discovered in classical decision making, appear
in decision problems that can be formulated as follows. One considers a set 
of outcome payoffs $X \equiv \{x_i: i = 1, 2, \ldots\}$, on which a probability
measure $p: X \rightarrow [0,1]$ is given. Over the payoff set, there are several
lotteries, or prospects, $\pi_j = \left \{ x_i \; , p_j(x_i): \; i = 1,2,\ldots
\right \}$, differing by the outcome probabilities. The payoff set is the domain 
of a utility function $u(x)$ that is a non decreasing concave function. The
expected utility of a lottery $\pi_j$ is defined as $U(\pi_j) = \sum_i u(x_i) 
p_j(x_i)$. A lottery $\pi_1$ is said to be preferable to $\pi_2$ if and only 
if $U(\pi_1) > U(\pi_2)$. And the lotteries are indifferent, when $U(\pi_1) = 
U(\pi_2)$. Suppose that the given data are such that, according to the classical
decision  making, a lottery $\pi_1$ is preferable or indifferent to $\pi_2$, 
that is, $U(\pi_1) \geq U(\pi_2)$, However, decision makers, when deciding 
between several lotteries under uncertainty and in the presence of risk, often
choose $\pi_2$, instead of $\pi_1$, thus, contradicting the prescription of 
utility theory.  

Because of the large number of paradoxes associated with classical decision 
making, there have been many attempts to change the expected utility approach, 
which has been classified as non-expected utility theories. There exists a number 
of such non-expected utility theories, among which we may mention some of the 
best known: prospect theory (Edwards, 1955; Kahneman and Tversky, 1979), 
weighted-utility theory (Karmarkar, 1978, 1979; Chew, 1983), regret theory 
(Loomes and Sugden, 1982), optimism-pessimism theory (Hey, 1984), dual-utility 
theory (Yaari, 1987), ordinal-independence theory (Green and Jullien, 1988), 
and quadratic-probability theory (Chew et al., 1991). More detailed information 
can be found in the review by Machina (2008).     

However, as has been shown by Safra and Sigal (2008), none of non-expected 
utility theories can explain all those paradoxes. The best that could be 
achieved is a kind of fitting for interpreting just one or, in the best case, 
a few paradoxes, while the other paradoxes remained unexplained. In addition,
spoiling the structure of expected utility theory results in the appearance 
of complications and inconsistencies. As has been concluded in the detailed 
analysis of Al-Najjar and Weinstein (2009), any variation of the classical 
expected utility theory ``ends up creating more paradoxes and inconsistencies 
than it resolves''.

Inconsistencies between utility theory and actual decision making are caused
by the assumption of the classical utility theory that decision makers are 
rational, while in reality their decisions are always influenced by subconscious 
feelings, various prejudices, and emotions. Taking these subjective features 
into account implies the necessity of resorting to a kind of behavioral decision
making (Simon, 1955).  

Biases, such as subconscious feelings, emotions, and other subjective effects,
are superimposed on the objective evaluation of utility, introducing some kind
of indeterminacy in decision-maker choices. The situation reminds that occurring
in quantum theory, where the physical reality is decorated by the probabilistic
description of phenomena. The idea that the functioning of the human brain 
could be described by the techniques of quantum theory has been advanced by one 
of the founders of quantum theory (Bohr, 1933;1958). Von Neumann, who is both a 
founding father of game theory and of expected utility theory on the one hand 
and the developer of the mathematical theory of quantum mechanics on the other 
hand, himself mentioned that the quantum theory of measurement can be interpreted 
as decision theory (von Neumann, 1955). 

The main difference between the classical and quantum techniques is the way 
of calculating the probability of events. As soon as one accepts the quantum 
way of defining the concept of probability, the latter generally becomes 
non-additive. And one immediately meets such quantum effects as interference 
and entanglement. The possibility of employing the techniques of quantum theory 
in several branches of sciences, that previously have been analyzed by 
classical means, is nowadays widely considered. As examples, we can mention 
quantum game theory (Eisert and Wilkens, 2000; Landsburg, 2004; Guo et al., 2008), 
quantum information processing and quantum computing 
(Williams and Clearwater, 1998; Nielsen and Chuang, 2000; Keyl, 2002).

After the works by Bohr (1933, 1958) and von Neumann (1955), there have been 
a number of discussions on the possibility of applying quantum rules for
characterizing the process of human decision making (Aerts and Aerts, 1994; 
Segal and Segal, 1998; Baaquie, 2004, 2009; Busemeyer et al., 2006; 
Bagarello, 2009; Lambert-Mogiliansky et al., 2009; Kitto, 2009; Pothos and 
Busemeyer, 2010; Leaw and Cheong, 2010; West and Grigolini, 2010; Zabaletta 
and Arizmendi, 2010). More references can be found in the recent review 
article (Yukalov and Sornette, 2009b). However, no general theory with 
quantitative predictive power has been suggested. This was the motivation for 
our introduction of a general quantum theory of decision making, based on 
the Hilbert-space functional analysis and von Neumann theory of quantum 
measurements (von Neumann, 1955), which could be applied to any possible 
situations (Yukalov and Sornette, 2008, 2009a,b,c, 2010a,b, 2011). Our approach 
is, to the best of our knowledge, the first theory using the mathematical 
formulation of quantum theory that allows for the {\it quantitative} treatment 
of different classical paradoxes in the frame of a single general scheme. 
Indeed, practically all paradoxes of classical decision making find their 
natural explanation in the frame of the Quantum Decision Theory 
(Yukalov and Sornette, 2008, 2009a,b,c, 2010a,b, 2011).

Our framework does not assume that decision makers are quantum objects. The 
techniques of quantum theory are employed just as a convenient mathematical 
tool allowing us to combine the notions of objective utility and subjective
biases. Actually, the sole thing we need is the theory of Hilbert spaces. 
Generally, it is worth stressing that the use of quantum techniques requires 
that neither brain nor consciousness would have anything to do with genuinely 
quantum systems. The techniques of quantum theory are used solely as a 
{\it convenient mathematical tool} and language to capture the properties 
associated with decision making. It is known that the description of any 
quantum system could be done as if it was a classical system, via the
introduction of the so-called contextual hidden variables. However, their 
number has to be infinite in order to capture the same level of elaboration 
as their quantum equivalent (Dakic et al., 2008), which makes unpractical the 
use of a classical equivalent description. Instead, quantum techniques are 
employed to describe systems in which interference and entanglement effects 
occur, because they are much simpler than to deal with a classical system 
having an infinite number of hidden unknown variables. Similarly, we use 
quantum techniques for decision theory in order to implicitly take into 
account the existence of many hidden variables in humans, such as emotions, 
subconscious feelings, and various biases. The existence of these hidden 
variables strongly influences decision making, as captured partially, for 
instance, by the notion of bounded rationality (Simon, 1955) and confirmed 
by numerous studies in Behavioral Economics, Behavioral Finance, Attention 
Economy, and Neuroeconomics (Cialdini, 2001; Loewenstein et al., 2008).

The standard setup displaying the paradoxes in classical decision making
corresponds to {\it individual} decision makers that take decisions without
consulting each other. However, in a number of experimental studies, it has 
been found that consultation sharply reduces errors in decision making. For
example, Cooper and Kagel (2005) and Blinder and Morgan (2005) find that 
groups consistently play more strategically than do individuals and generate
positive synergies in more difficult games. Charness et al. (2007a,b) show 
that group membership affects individual choices in strategic games. Charness 
and Rabin (2002) and Chen and Li (2009) investigate the minimal-group paradigm 
and find a substantial increase in charity concerns and social-welfare-maximizing
actions when participants are matched with in-group members. It was found that 
the errors in the famous disjunction effect and conjunction fallacy strongly
attenuate when group members get information by learning from their experience
(K\"{u}hberger et al., 2001) or exchange information by consulting (Charness 
et al., 2010). Groups usually perform better than individuals at quantitative 
judgment tasks (Sung and Choi, 2012; Schultze et al., 2012). 

Explaining these attenuation effects, caused by information transfer through
the interactions between decision makers, requires extending the theory from 
isolated individuals to human beings who are part of a society within which
they interact and exchange information. It is the aim of the present paper to 
generalize the QDT approach to the case of decision makers who interact within 
a group or society. The information received from the society influences the 
decisions. This leads to a natural explanation of the error attenuation effect, 
as compared with the paradoxes existing for decisions without within-group 
consultations.   

In the next Section 3, we present the generalization of QDT for a decision 
maker who is not a separate individual, but a member of a society. In the 
following Section 4, we show how the additional information, received by the 
decision maker through interactions with the surrounding society, leads to a 
decrease of errors compared with classical decision making. We discuss the 
experiments by Charness et al. (2010) and explain why the initial error in the 
conjunction fallacy diminishes with the received information.

\section{Social decision makers}

Let us consider a society defined as a collective of several agents. Each 
agent is a decision maker, whose decisions are influenced by other members of 
the society. A decision maker aims at choosing between several admissible 
choices, called lotteries or prospects. Note that a decision maker can also 
be represented by a group jointly making a choice through a group decision making
(Xu, 2011; Tapia et al, 2012). In that case, the society is understood as a
collection of several such groups, each acting as a separate decision maker.
Our aim is to study how the choice of individual decision makers is influenced
by their interactions in a society, leading to the exchange of information. 

It is necesssary to mention that decision making in a society has been 
considered by invoking the techniques of statistical mechanics, following the 
Brock-Durlauf (1999, 2001) approach. Statistical models are usually formulated
as some variants of the Ising model, where a node is associated with a single 
decision maker deliberating on a binary choice ``yes-or-now". In that sense,
an undividual decision maker possesses a classical bit of information and acts
as a subject obeying classical rules (Durlauf, 1999; Brock, 2001; Contucci, 2008;
Barra, 2010; Barra, 2012).               

In the approach of Quantum Decision Theory (Yukalov and Sornette, 2008, 2009a,b,c, 
2010a,b, 2011), a single decision maker enjoys quantum bits of information. 
This becomes especially important and makes a principal difference with the 
classical way of decision making when the considered prospects are composite.  
Then, even in the case of a binary ``yes-no" problem, in the presence of uncertainty, 
the available information is characterized by {\it quantum qubits}, which leads 
to the appearance of interference effects. Such interference effects do not occcur 
for a single decision maker dealing with a simple binary choice, if no composite 
events are present. The latter situation happens in classical societies described
by statistical models. In such models, uncertainty is caused by the fact that 
an individual decision maker does not know what are the decisions of other members
of the society. Contrary to this, in the quantum approach, even a single decision 
maker experiences uncertainty, without the presence of any other members. This is 
what principally distinguishes our approach from the classical statistical models. 
In our case, there exist two types of uncertainty: purely quantum, occuring even 
for a single decision maker, and statistical, caused by the existence of other
members of the society.      

In our previous papers (Yukalov and Sornette, 2008, 2009a,b,c, 2010a,b, 2011), 
we have considered separate decision makers. For each prospect, we associate a 
vector in a Hilbert space. But now, in addition to the space of mind for a given 
separate decision maker, there exists the decision space of the society as a whole. 
Below, we give a generalization of QDT for social decision makers. All mathematical definitions and their relation to real-life situations have been thoroughly 
explained in our previous publications, and we do not think it would be appropriate 
to essentially extend the present paper by repeating all those technical details. 
Rigorous mathematical techniques for defining the quantum probabilities of composite
prospects can be found in (Yukalov and Sornette, 2013). In the following 
sections, we give only the necessary mathematical minimum for correctly describing 
and justifying the use of the prospect probabilities. The readers, who are not 
accustomed to the techniques of Hilbert-space functional analysis, can omit the mathematical explanations and can jump to the end of this section, just accepting 
the form of the prospect probabilities and the rules they satisfy.

\subsection{Decision spaces}

Let an agent $A$ be a member of a society. Assume that, for this agent, there 
exists a set of elementary prospects that are represented by a set 
of vectors $\{|n \rgl\}$. The elementary-prospect vectors are orthonormalized, 
so that the scalar product $\lgl m|n\rgl = \delta_{mn}$ is a Kronecker delta. 
The orthogonality of the elementary prospects means that they are independent 
and not compatible, so that only one of them can be realized. The elementary 
prospects represent the variety of separate possible actions that could be 
accomplished by the decision maker. For instance, such possible actions could is 
be represented by choosing whom to marry and what job to accept. The detailed
life examples can be found in Yukalov and Sornette (2009a,b, 2010b, 2011).    
Strictly speaking, the prospects are assumed to be well defined. This 
distinguishes our consideration from the case of imprecise knowledge of prospects
and imprecise available information, which requires the use of the fuzzy decision
theory (Aliev et al. 2012). 

The space of mind of a decision maker, by definition, is a closed linear 
envelope
\be
\label{1}
\cH_A \equiv {\rm Span} \{ | n \rgl \}
\ee
spanning all admissible elementary prospects. Then the elements of the space 
of mind represent all admissible combinations of various actions. Similarly, 
such a space of mind can be constructed for each member of the society, the 
states of mind of two distinct individuals being in general different. Let 
the space of mind for all members of the society, except the agent $A$, be 
denoted as $\mathcal{H}_B$. Then the total decision space of the whole society 
is the tensor product
\be
\label{2}
 \cH_{AB} \equiv \cH_A \bigotimes \cH_B \;  .
\ee
This is a Hilbert space, where a scalar product is defined. The space 
$\mathcal{H}_B$ can also be presented as a tensor product of the individual 
spaces of all other society members. 

The elementary prospects serve as a basis for constructing the Hilbert space 
of mind. But they are not necessarily the prospects a decision maker is 
evaluating. They just enumerate all admissible possibilities. But, generally,
a decision maker deliberates choosing not between the elementary prospects, 
but between combinations of these. For instance, one may decide whether to 
accept a particular job, under the condition of marrying either one or 
another person (see details in our cited papers).

\subsection{Prospect states}

The decision maker $A$ considers a set of prospects
\be
\label{3}
 \cL = \{ \pi_j : \; j = 1,2, \ldots N \} \;  .
\ee
Each prospect $\pi_j$ is put into correspondence to a vector $|\pi_j\rgl$,
called the prospect state, in the space of mind $\mathcal{H}_A$. The prospects 
of $\mathcal{L}$ are, generally, composite objects composed of several 
elementary prospects. Many concrete examples are given in the published papers 
(Yukalov and Sornette, 2009a,b, 2010b, 2011).

Being an element of the space $\mathcal{H}_A$, a prospect state can be
represented as an expansion over the elementary prospects,
\be
\label{4}
  | \; \pi_j \; \rgl = 
\sum_n \; \lgl\; n \;| \; \pi_j \; \rgl \; | \; n \;\rgl \; . 
\ee
The prospect states are not assumed to be either orthogonal or normalized,
so that the scalar product
\be
\label{5}
\lgl \;\pi_i \;| \;\pi_j\; \rgl = \sum_n \; \lgl \;\pi_i \;| \;n \;\rgl 
\lgl \;n \;| \; \pi_j \;\rgl
\ee
is not a Kronecker delta. The prospects states are not orthogonal with each 
other, since they are not necessarily incompatible, but can interfere and 
entangle with each other. And the appropriate normalization condition will 
be imposed later.  
    
The prospects are the targets of the decision maker in the sense that he/she 
chooses which of them to prefer. As far as one can compare the prospects by 
qualifying them as more or less preferable, the set $\mathcal{L}$ of these 
prospects $\pi_j$ should be ordered, forming a complete transitive lattice. 
The ordering procedure will be given below. The aim of decision making is 
to find out which of the prospects is the most favorable. 

There can exist two types of setups. One is when a number of agents choose 
between the given prospects. Another type is when a single decision maker 
takes decisions in a repetitive manner, for instance taking decisions several 
times. These two cases are treated similarly.

\subsection{Prospect operators}

To each prospect $\pi_j$, with a vector state $|\pi_j \rgl$ in the Hilbert space 
of mind $\mathcal{H}_A$, there corresponds the prospect operator
\be
\label{6}
\hat P(\pi_j) \equiv | \; \pi_j \; \rgl \lgl \; \pi_j \; | \; .
\ee
By this definition, the prospect operators are self-adjoint. These operators, 
generally, are not projectors, as far as they are not necessarily idempotent,
$$
 \hat P^2(\pi_j) = \lgl \;\pi_j \;| \; \pi_j \; \rgl \hat P(\pi_j) \; ,
$$
which follows from the fact that the prospect states, generally, are not 
normalized. The prospect operators are not commutative, since the expressions
$$
 \hat P(\pi_i) \hat P(\pi_j) = \lgl \;\pi_i \;| \;\pi_j \; \rgl |\; \pi_i \;\rgl 
\lgl \; \pi_j \; | \;  , \qquad
 \hat P(\pi_j) \hat P(\pi_i) = \lgl \;\pi_j \;| \;\pi_i \; \rgl |\; \pi_j \;\rgl 
\lgl \; \pi_i \; | \;  ,
$$
differing by the order of operators, are not equivalent. The noncommutativity
of the prospect operators represents the noncommutativity of decisions in real 
life (Yukalov and Sornette, 2009a,b, 2010b, 2011).

The collection $\{\hat P(\pi_j)\}$ of the prospect operators is analogous
to the algebra of local observables in quantum theory. In the latter, as 
is known, not each product of local observables is, strictly speaking, an 
observable. But it is always possible to define symmetrized products so that 
the collection of local observables would form an algebra. In the same way
as for the operators of local observables in quantum theory, we can consider
the family $\{\hat P(\pi_j)\}$ of prospect operators as an algebra of 
observables in QDT.

Note that we use here the standard terminology related to operator algebras 
in Hilbert spaces (Neumann, 1955). The operator of an observable is not, of 
course, an observable quantity by itself, but it corresponds to such a 
quantity that is obtained by defining the operator expected value.

\subsection{Prospect probabilities}

QDT is a probabilistic theory, whose observable quantities are the prospect
probabilities. These prospect probabilities are defined as the expected 
values, that is, averages of the prospect operators. In that sense, the 
prospect probabilities play the role of the observable quantities 
(Yukalov and Sornette, 2008). In our previous papers, the averages were 
defined with respect to a given {\it strategic state} $|\psi\rgl$ 
characterizing the decision maker. Such a procedure corresponds to the 
averaging over a prescribed pure state, which assumes that the considered 
decision maker is an individual, not interacting with any surrounding. But 
when considering a decision maker in a society, which he/she interacts with, 
such a decision maker cannot be characterized by a pure state. 

The society as a whole could be described by a pure wave function, with the 
decision maker being a part of the society, which would then lead to the 
necessity of characterizing this decision maker by a statistical operator. 
This is in a direct analogy with treating subsystems of large systems by 
density matrices (Coleman and Yukalov, 2000). The statistical operator 
characterizes the state of the system as a whole, weighting the admissible 
microscopic states, in that sense being analogous to the probability 
distribution of classical probability theory.   

Moreover, we could describe the society by a wave function only if we would 
assume that the society is completely isolated from its surrounding. But 
such an assumption is certainly unreasonable, since there are no absolutely 
isolated societies. Again, this is completely equivalent to the absence of
absolutely isolated finite quantum systems (Yukalov, 2002, 2003a,b). Thus,
the most general way of describing the society state is by a statistical 
operator.

In the present case, the society state, including the considered decision 
maker, is to be characterized by a statistical operator $\hat{\rho}_{AB}$ 
that is a positive operator on $\mathcal{H}_{AB}$, normalized as
\be
\label{7}
{\rm Tr}_{AB} \hat\rho_{AB} = 1 \;   ,
\ee
with the trace operation being performed over $\mathcal{H}_{AB}$. The observable 
quantities are to be defined by the expectation values over the statistical 
state. Therefore the prospect probabilities are given by the averages
\be
\label{8}
  p(\pi_j) \equiv {\rm Tr}_{AB} \hat\rho_{AB} \hat P(\pi_j) \; .
\ee

The prospect operators act on the space of mind $\mathcal{H}_A$ of the
decision maker. Hence the above average can be represented as
\be
\label{9}
 p(\pi_j) \equiv {\rm Tr}_{A} \hat\rho_{A} \hat P(\pi_j) \;  ,
\ee
where the trace is over $\mathcal{H}_A$ and the reduced statistical operator is 
\be
\label{10}
 \hat\rho_A \equiv {\rm Tr}_B\hat\rho_{AB} \;  .
\ee
This operator characterizes the decision maker in the society. The
reduction to the previous situation of a single separated decision maker,
as considered in our previous papers, would correspond to the 
representation of the statistical operator $\hat{\rho}_A$ in the pure form 
$|\psi \rgl \lgl \psi|$, with the state $|\psi \rgl$ being the decision maker
strategic state. But, generally, the statistical operator $\hat{\rho}_A$
cannot be represented in such a factor form, since the decision maker state
is entangled with that of the society.

Introducing the matrix elements over the elementary-prospect basis for the
statistical operator 
\be
\label{11}
\rho_{mn} \equiv \lgl \; m \; | \; \hat\rho_A\; |\; n\; \rgl
\ee
and for the prospect operators
\be
\label{12}
  P_{mn}(\pi_j) \equiv \lgl \; m \;|\; \hat P(\pi_j)\; |\; n \;\rgl =
\lgl \; m \; |\; \pi_j \; \rgl \lgl \; \pi_j\; |\; n \;\rgl 
\ee
makes it possible to rewrite the prospect probabilities as
\be
\label{13}
 p(\pi_j) =\sum_{mn} \rho_{mn} P_{nm}(\pi_j) \;  .
\ee
      
To really represent probabilities, the above quantities are to be normalized
so that
\be
\label{14}
 \sum_{j=1}^N \; p(\pi_j) = 1 \;  .
\ee
Since the statistical operator, by definition, is a positive operator, we have
\be
\label{15}
 0 \leq p(\pi_j) \leq 1 \;  .
\ee
This defines the collection $\{ p(\pi_j) \}$ as a probability measure. The most 
favorable prospect corresponds to the largest of the probabilities.

Let us introduce the {\it utility factor}
\be
\label{16}
f(\pi_j) \equiv \sum_n \rho_{nn} P_{nn}(\pi_j)
\ee
and the {\it attraction factor} 
\be
\label{17}
  q(\pi_j) \equiv \sum_{m\neq n} \rho_{mn} P_{nm}(\pi_j) \; ,
\ee
whose meanings will be explained below. Then, separating the diagonal and 
non-diagonal terms in the sum over $m$ and $n$, we obtain the probability 
of a prospect $\pi_j$ as the sum
\be
\label{18}
 p(\pi_j) = f(\pi_j) + q(\pi_j)  
\ee
of the above two factors.   

Though some intermediate steps of the theory might look a bit complicated, the 
final result is rather simple and can be straightforwardly used in practice,
provided the way of evaluating the utility and attraction factors are known.

\subsection{Utility factors and correspondence with classical utility theory}

As is known (Neumann, 1955), the expectation values of observables in quantum 
theory can be separated in two terms, one having a diagonal representation 
over the chosen basis and another being off-diagonal in this representation. 
The diagonal part corresponds to the classical value of the observable, while 
the off-diagonal part characterizes purely quantum effects caused by interference. 
The same holds in our case, where the prospect probability (\ref{8}) is defined 
as the expectation value of the prospect operator. The diagonal part is the 
utility factor (\ref{16}) describing the weight of the prospect calculated 
classically. To be defined as a weight, the set of these factors is to be 
normalized as
\be
\label{19}
 \sum_{j=1}^N f(\pi_j) = 1 \; ,
\ee
from where one has
\be
\label{20}
 0 \leq f(\pi_j) \leq 1 \;  ,
\ee
since, by definition (\ref{16}), the factor is non-negative.

In classical decision theory, the choice of a decision maker is based on 
the notion of expected utility. One considers a set of measurable payoffs 
$\{x_i\}$ associated with the related probabilities $p_j(x_i)$ whose family 
forms a probability measure with the standard properties
$$
 \sum_i \; p_j(x_i) = 1 \; , \qquad 0 \leq p_j(x_i) \leq 1 \;  .
$$
A prospect $\pi_j$ is represented by a lottery
\be
\label{21}
 \pi_j \equiv \{ x_i \; , p_j(x_i): \; i = 1,2, \ldots \} \; .
\ee
Linear combinations of lotteries are defined as
$$
\sum_j \lbd_j \pi_j = \left \{ x_i \; , \;\; 
\sum_j \lbd_j p_j(x_i) \right \} \;   ,
$$
with the constants $\lambda_j$ such that
$$
 \sum_j \lbd_j = 1 \; , \qquad 0 \leq \lbd_j \leq 1 \;  .
$$

Introducing a utility function $u(x)$, which is defined as a non-decreasing
and concave function, one constructs the expected utility
\be
\label{22}
 U(\pi_j) = \sum_i u(x_i) p_j(x_i) \;  .
\ee

As quantum decision theory is a more general theory than classical utility 
theory, it is desirable to formulate a correspondence such that the 
predictions of quantum decision theory would reduce to those of classical 
decision theory in some limit to be defined. For this, since quantum 
decision theory is intrinsically probabilistic, while classical decision 
theory is deterministic (the prospect with the largest expected utility 
is assumed to be chosen with certainty), we need to first generalize the
classical utility theory to endow it with a probabilistic skin, which 
itself could be reduced to the deterministic form of utility theory under 
appropriate conditions. Then, the correspondence between quantum decision 
theory and classical utility theory could be formulated via the 
probabilistic extension of the latter. In essence, when decoherence occurs, 
the attraction factors in quantum decision theory tend to zero and the 
prospect probabilities should tend to the utility factor of the probabilistic 
version of classical decision theory. The utility factor is interpreted 
as the probability of having the given utility. Note that several variants 
of such a probabilistic reformulation of classical utility theory have 
been considered (Luce, 1958; Marschinski et al., 2007; Cadogan, 2011) by 
postulating the Boltzmann distribution of prospects. The latter, however, 
does not satisfy the necessary boundary condition requiring that the 
prospect of zero utility should have zero weight:
\be
\label{A1}
 f(\pi_j) \ra 0 \qquad ( U(\pi_j) \ra 0 ) \;  .
\ee
Here, we show how the correct form of the utility factor can be derived.

The most general way of deriving distributions is suggested by the 
principle of minimal information allowing one to define the most accurate
distribution form, given a number of known constraints or facts, under 
the condition of assuming the minimum amount of additional structures 
in the absence of other information on the system. The relevant tool is 
thus to minimize the Kullback-Leibler information function in the presence 
of the constraints. The necessary first step in this procedure is the 
definition of a representative statistical ensemble, taking into account 
all conditions that uniquely characterize the considered statistical 
system (Yukalov, 2007).

In the probabilistic formulation of utility theory, the prospects $\pi_j$ 
are treated as random events. Then, the prospect lattice (\ref{3}) plays
the role of the field of these random events. As a consequence, the expected 
utility $U(\pi_j)$ is also interpreted as a random quantity, which, for 
concreteness, is assumed to be non-negative. Therefore, there should exist 
a probability measure describing the distribution of the expected utilities.

The corresponding probability $f(\pi_j)$ of a prospect $\pi_j$ is to be
normalized according to (\ref{19}). We also impose that the average of the 
random expected utilities should exit, giving the total utility
\be
\label{A2}
 \sum_j U(\pi_j) f(\pi_j) = U \;  . 
\ee
Imposing that the average utility exists as a given value $U$ means that 
the decision maker is operating under the hope of keeping his/her level 
of satisfaction constant. This is close in spirit to the notion of 
``satisficing'' introduced by Simon (1956), in which a person stops 
searching and optimizing when a sufficient level of utility is reached.

The prospect that yields the maximal utility,
\be
\label{A3}
 U(\pi_{max}) \equiv \max_j U(\pi_j) \;  ,  
\ee
plays a special role for the decision maker. In the deterministic version 
of classical utility theory, it is the chosen prospect, with certainty. 
This suggests to normalize the utilities of each prospect by this value 
taken as a reference. We thus introduce a kind of likelihood function
\be
\label{A4}
f_0(\pi_j) = C \; \frac{U(\pi_j)}{U(\pi_{max} ) } \;,
\ee
where $C$ is a normalization coefficient. This function quantifies the 
relative value of each expected utility in units of the maximal utility 
value taken as the natural reference. Here, we capture the fact that 
typical decision makers ponder their options in relative terms.

Then, the utility factor $f(\pi_j)$ should be such as obeying the 
conditions (\ref{19}) and (\ref{A2}) and being as close as possible to 
the reference likelihood function $f_0(\pi_j)$ defined by expression 
(\ref{A4}). For this, we use the distance measure or relative information 
between $f(\pi_j)$ and $f_0(\pi_j)$ 
\be
\label{A5}
\sum_j f(\pi_j) \ln \; \frac{f(\pi_j)}{f_0(\pi_j)} \; , 
\ee
introduced by Kullback and Leibler (1951, 1959). Note that
expression (\ref{A5}) tends to zero as $f(\pi_j) \to f_0(\pi_j)$.

The representative statistical ensemble is the pair $\{{\cal L}, f(\pi_j)\}$ 
of the field of events, that is, of random prospects, and of the prospect 
distribution $f(\pi_j)$, under conditions (\ref{19}) and (\ref{A2}), with 
the Kullback-Leibler relative information (\ref{A5}). The corresponding 
information functional reads as
$$
I[f(\pi_j) ] = \sum_j f(\pi_j) \; \ln\; \frac{f(\pi_j)}{f_0(\pi_j) } \; +
$$
\be
\label{A6}
+ \lbd \left [ \sum_j f(\pi_j) - 1 \right ] -
\bt \left [ \sum_j f(\pi_j) U(\pi_j) - U \right ] \; , 
\ee
where $\lambda$ and $\beta$ are Lagrange multipliers.

Minimizing the information functional (\ref{A6}) yields the distribution
\be
\label{A7}
f(\pi_j) = \frac{U(\pi_j)}{Z} \; \exp\{ \bt U(\pi_j) \} \; ,
\ee
with the normalization factor
$$
Z = \sum_j  U(\pi_j) \exp\{ \bt U(\pi_j) \} \; . 
$$
The parameter $\beta$ can be called the {\it confidence}, or {\it belief},
or {\it certainty parameter}, since it characterizes how confident is the 
choice of the prospect lattice. Requiring that $f(\pi_j)$ would increase 
together with $U(\pi_j)$ implies that the confidence parameter is to be 
non-negative: $\beta \geq 0$. We stress that the utility factor is 
a normalized function of the expected utility, but not the utility itself. 
This function increases together with the utility. 

In the limiting case of absolute certainty, when $\beta \ra \infty$, we
return to the completely deterministic choice of the prospect with the 
maximal expected utility, as specified by classical utility theory:
\begin{eqnarray}
\nonumber
f(\pi_j) = \left \{ \begin{array}{ll} 
1 , & ~ \pi_j = \pi_{max} \\
0 , & ~ \pi_j \neq \pi_{max} \end{array}
\qquad (\bt \ra \infty) \; . \right.
\end{eqnarray}
And in the case of the completely uncertain choice of prospects, when 
$\beta = 0$, we come to the form of the utility factor 
\be
\label{23}
f(\pi_j) = \frac{U(\pi_j)}{\sum_j U(\pi_j) } \; .
\ee
It is worth emphasizing that the utility factor (\ref{A7}) satisfies the 
necessary limiting condition (\ref{A1}), telling that the weight
(or probability to be chosen) of a prospect having no utility is zero. 

The parameter $\beta$ constitutes an important ingredient in the 
formulation of a probabilistic version of utility theory. We expect it 
to be individual-dependent. What quantum decision theory adds is the 
existence of coherence and interference between prospects, which are 
quantified by the attraction factors.

In summary, quantum decision theory reduces to the probabilistic version 
of utility theory specified in terms of the utility factors (\ref{A7})
in the absence of coherence and interference effects. These utility 
factors (\ref{A7}) give the probability that a prospect, with a given 
expected utility, is chosen. And our probabilistic version of utility 
theory reduces to the classical deterministic utility theory when the 
belief factor $\beta$ tends to infinity. This concludes the construction
of the correspondence between classical utility theory and quantum 
decision theory.

\subsection{Attraction factors}

The off-diagonal term in the expectation value (\ref{9}) is the attraction
factor (\ref{17}) representing quantum interference, or coherence, effects.
In QDT, the attraction factor is a contextual object describing subconscious 
feelings, emotions, and biases, playing the role of hidden variables. 
Despite their contextuality, the attraction factors satisfy some general
properties that make possible their quantitative evaluation.  

In view of normalizations (\ref{14}) and (\ref{19}), the attraction factors 
satisfy the {\it alternation property}, such that the sum
\be
\label{24}
\sum_{j = 1}^N \; q(\pi_j) = 0 \; 
\ee
over the prospect lattice $\mathcal{L}$ is always zero, and the values 
of the attraction factor are in the range
\be
\label{25}
-1 \leq q(\pi_j) \leq 1 \; .
\ee
In addition, the average absolute value of the attraction factor is estimated 
(Yukalov and Sornette, 2009b, 2011) by the {\it quarter law}
\be
\label{26}
\frac{1}{N} \; \sum_{j=1}^N \; |\; q(\pi_j) \; | = \frac{1}{4} \;   .
\ee
These properties allow us to {\it quantitatively} define the prospect 
probabilities (\ref{18}).

We may note that the attraction factor exists only for composite prospects,
composed of several actions, while for elementary prospects this term is zero.
This is easy to show as follows. Let $e_j$ be an elementary prospect 
corresponding to a state $|j \rgl$, hence $\lgl n|j \rgl = \delta_{nj}$.
The related prospect operator $\hat{P}(e_j)$ is defined in Eq. (\ref{6}). 
Then the prospect probability (\ref{13}) reduces to 
$$
 p(e_j) =\sum_{mn} \rho_{mn} \dlt_{mj} \dlt_{nj} = \rho_{jj} \;  ,
$$
and the attraction factor is zero:
$$
 q(e_j) = 0\;  .
$$

In this way, there exists a direct general relation between Quantum 
Decision Theory and classical decision theory, based on the maximization 
of expected utility. Classical decision theory is retrieved when the 
attraction factor is zero. The form of the utility factor (\ref{23}) shows
that, in this situation, maximizing the expected utility is equivalent
to maximizing the utility factor. Thus, classical decision theory is a 
particular case of the more general QDT in the case when only objective 
information on the decision utility is taken into account, while subjective 
sides, such as biases, emotions, and subconscious feelings play no role. 
The latter variables do play a very important role in decision making 
performed in many important and practical situations. Our approach takes 
into account both the objective utility of considered prospects as well as 
their subjective attractiveness for the decision maker.      

Let us briefly summarize. As we said, the attraction factor in QDT appears 
naturally in order to account for subconscious feelings, emotions, and biases. 
Despite the fact that the attraction factor is contextual, it satisfies three 
pivotal general properties: (i) an attraction factor varies in the interval 
$[-1, 1]$; (ii) the sum of all attraction factors over the lattice of considered 
prospects is zero; (iii) the average absolute value of an attraction factor 
is 0.25. These properties make it possible to give a {\it quantitative} 
evaluation of prospect probabilities and, thus, to develop a practical way 
of applying QDT to realistic problems of decision making.

\subsection{Prospect ordering}

Since the prospect probability (\ref{18}) consists of two terms, we should 
consider both of them, when comparing the probabilities of different prospects.
That is, we have to compare the usefulness as well as attractiveness of the
prospects.

The usefulness of prospects is measured by the utility factor. The prospect 
$\pi_1$ is more useful than $\pi_2$, when 
\be
\label{27}
f(\pi_1) > f(\pi_2) \; . 
\ee
The prospects $\pi_1$ and $\pi_2$ are equally useful, if
\be
\label{28}
 f(\pi_1) = f(\pi_2) \;  .
\ee
And the prospect $\pi_1$ is not less useful (more useful or equally useful) 
than $\pi_2$, if
\be
\label{29}
 f(\pi_1) \geq f(\pi_2) \;  .
\ee

The attractiveness of prospects is characterized by their attraction factors. 
The prospect $\pi_1$ is more attractive than $\pi_2$, if 
\be 
\label{30}
q(\pi_1) > q(\pi_2) \; .
\ee
The prospects $\pi_1$ and $\pi_2$ are equally attractive, when
\be
\label{31}
 q(\pi_1) = q(\pi_2) \;  .
\ee
And the prospect $\pi_1$ is not less attractive (more attractive or equally 
attractive) than $\pi_2$, when
\be
\label{32}
 q(\pi_1) \geq q(\pi_2) \;  .
\ee

The comparison between the attractiveness of prospects can be done on the 
basis of the {\it aversion to uncertainty and risk} or {\it ambiguity aversion}
(Rothschild and Stiglitz, 1970; Gollier, 2001; Sornette, 2003; Malvergne
and Sornette, 2006; Abdellaoui et al., 2011a; 2011b; Yukalov and Sornette, 2011).

For example, a prospect is more attractive when:
\begin{itemize}
\item[(i)] it provides more certain gain (more uncertain loss).
\item[(ii)] it promotes to be active under certainty (passive under uncertainty).   
\end{itemize}

The total evaluation of prospects that finally influences the decision maker 
choice is based on the prospect probabilities. The prospect $\pi_1$ is 
preferable to $\pi_2$, if
\be
\label{33}
 p(\pi_1) > p(\pi_2) \;  .
\ee
The prospects $\pi_1$ and $\pi_2$ are indifferent, when
\be
\label{34}
 p(\pi_1) = p(\pi_2) \;  .
\ee
And the prospect $\pi_1$ is preferable or indifferent to $\pi_2$, if
\be
\label{35}
 p(\pi_1) \geq p(\pi_2) \;  .
\ee

The classification of prospects of a set $\mathcal{L}$ as more or less 
preferable establishes an order in $\mathcal{L}$ making this ordered set
a {\it lattice}. Among all prospects, there exists the least preferable 
prospect with the minimal probability, and the most preferable prospect
with the largest probability. Hence, the prospect lattice $\mathcal{L}$ 
is {\it complete}. The lattice is also {\it transitive} since, if $\pi_1$ 
is preferable to $\pi_2$, with $\pi_2$ being preferable to $\pi_3$, then
$\pi_1$ is preferable to $\pi_3$.
 
Decision makers may choose the most preferable prospect, whose probability  
is the largest. Such a prospect is called {\it optimal}. The prospect $\pi_*$
is optimal if and only if
\be
\label{36}
 p(\pi_*) =\max_j p(\pi_j) \;  .
\ee 

In the presence of two criteria characterizing each prospect, a given prospect 
can be more useful, while being less attractive, or vice-versa. As a consequence, 
there are situations where the ordering of classical utility theory is inverted, 
so that the less useful though more attractive prospect is preferred, having the 
largest probability. This important fact can be formalized by the following 
statement.

\vskip 2mm

{\bf Proposition 1}. {\it The prospect $\pi_1$ is preferable to $\pi_2$ if and 
only if}
\be
\label{37}
 f(\pi_1) - f(\pi_2) > q(\pi_2) - q(\pi_1) \;  .
\ee

{\it Proof}: It follows from the comparison of the prospect probabilities 
(\ref{18}) for $\pi_1$ and $\pi_2$.

\vskip 2mm

This inequality provides an explanation for the appearance of paradoxes in 
classical decision making as resulting from the role of the attraction factor 
representing the interference between prospects caused by behavioral biases. 
It is remarkable that this simple idea seems to be sufficient to remove the 
empirical paradoxes and make QDT consistent with the decisions made by real 
human beings. The existence of the attraction factor is due to the presence 
of risk and uncertainty associated with the choices to be made.

\subsection{Binary lattice}

A situation that is very often considered in empirical research consists
in choosing between two prospects, which corresponds to a binary lattice
\be
\label{38}
 \cL = \{ \pi_1 \; , \pi_2 \} \;  .
\ee
This case is sufficient to clearly illustrate the above general 
considerations. 

For a binary lattice, we have
\be
\label{39}
p(\pi_1) = f(\pi_1) + q(\pi_1) \; , \qquad  
p(\pi_2) = f(\pi_2) + q(\pi_2) \;  .
\ee
The normalization (\ref{19}) reads as
\be
\label{40}
 f(\pi_1) + f(\pi_2) = 1\;  ,
\ee
and the alternation property (\ref{24}) becomes
\be
\label{41}
 q(\pi_1) + q(\pi_2) = 0\;  .
\ee

If the considered two prospects are equally attractive, which implies
$q(\pi_1) = q(\pi_2)$, then, according to (\ref{41}), we get 
$q(\pi_1) = q(\pi_2) = 0$. Therefore, the prospect probabilities 
coincide with their utility factors, $p(\pi_1) = f(\pi_1)$ and
$p(\pi_2) = f(\pi_2)$. In such a situation, we return to the standard 
decision making recipe based on the comparison between the prospect 
utilities.

But when the prospects are not equally attractive, say $\pi_1$ is more 
attractive than $\pi_2$, that is, $q(\pi_1) > q(\pi_2)$, then the 
alternation property (\ref{41}) yields
$$
 q(\pi_1) = - q(\pi_2) > 0\;  .
$$
This allows one to make accurate predictions of the choice
of real human beings who have to choose an optimal prospect.

\subsection{Individual decisions}

Suppose that a decision maker has to choose between several prospects.
Let he/she be assumed to make a decision sufficiently quickly, with
no consultations with other members of society, and without getting 
additional information from other sources. This kind of decision making
can be termed {\it individual} or {\it spontaneous}. Such a setup is 
typical of the majority of experimental observations, where different 
paradoxes have been documented.  

In the case of this spontaneous decision making, it is possible to
quantitatively predict typical decisions and, respectively, to explain
the occurrence of characteristic paradoxes. This can be done as follows.
Let us consider a binary lattice of prospects. Assume that, according to
the risk-uncertainty aversion formulated above, the prospect $\pi_1$
is more attractive than $\pi_2$, hence 
$$
q(\pi_1) > q(\pi_2)\;   .
$$

It is possible to estimate the attraction factors by their mean values,
as explained above, evaluating $q(\pi_1)$ as equal to $1/4$ and $q(\pi_2)$ 
as given by $-1/4$. At the same time, the probability belongs to the interval 
$[0,1]$. To take this into account, it is convenient to invoke the function, 
called {\it retract}, such that 
\begin{eqnarray}
\nonumber
{\rm Ret}_{[a,b]}\{ z \} =\left \{
\begin{array}{ll}
a \; , & ~ z \leq a \\
z \; , & ~a < z < b \\
b \; , & ~ z \geq b 
\end{array} \; . \right.
\end{eqnarray}
Then the prospect probabilities (\ref{39}) can be represented as
\be
\label{42} 
 p(\pi_1) = {\rm Ret}_{[0,1]} \left \{ f(\pi_1) + 
\frac{1}{4} \right \} \; , 
\qquad
p(\pi_2) = {\rm Ret}_{[0,1]} \left \{ f(\pi_2) - 
\frac{1}{4} \right \} \; .
\ee
Since, the utility factors are calculated by means of formula (\ref{23}),
one gets a quantitative estimate for the prospect probabilities, which 
makes it possible to choose the preferable prospect.

\vskip 2mm
{\bf Proposition 2}. {\it Let the prospect $\pi_1$ from a binary prospect 
lattice be more attractive than $\pi_2$ and let the prospect probabilities 
be evaluated by expressions (\ref{42}), then $\pi_1$ is preferable over 
$\pi_2$ when the utility factor of $\pi_1$ is such that}
\be
\label{43}
 f(\pi_1) > \frac{1}{4} \qquad ( \pi_1 > \pi_2 ) \;  .
\ee
{\it Respectively, the prospects are indifferent, if $f(\pi_1) = 1/4$ and 
the prospect $\pi_2$ is preferable, if $f(\pi_1) < 1/4$}.

\vskip 2mm

{\it Proof}: It follows from expressions (\ref{42}) and the condition that
the prospect $\pi_1$ is more attractive than $\pi_2$, so that  
$q(\pi_1) > q(\pi_2)$.

\subsection{Comparison with empirical observations}

Strictly speaking, being defined to reflect subjective factors embodying 
subconscious feelings, emotions, and biases, the attraction factors are 
contextual. This means that their values can be different for different 
decision makers. Moreover, they can be different for the same decision 
maker at different times. These features seem to be natural when one 
keeps in mind that we are describing real humans, whose decisions are usually 
different, even under identical conditions. It is also known that the same 
decision maker can vary his/her decisions at different times and under 
different circumstances. However, focusing solely on the contextual character 
of the interference terms, gives the wrong impression of a lack of predictive 
power of the approach, which would make it rather meaningless.

Fortunately, there is a way around the problem of contextuality, based on the 
fact that QDT has been constructed as a probabilistic theory, with the 
probabilities interpreted in the frequentist sense. This is equivalent to 
saying that QDT is a theory of the aggregate behavior of a population. In 
other words, the predictions of the theory are statistical statements 
concerning the population of individualistic behaviors, namely, QDT provides 
the probability for a given individual to take this or that decision,
interpreted in the sense of the fraction of individuals taking these decisions.

The prospect probabilities, calculated in the frame of QDT, can be compared
with the results of experimental tests. In experiments, one usually 
interrogates a pool of $M$ decision makers, asking them to choose a 
prospect from the given prospect set $\{\pi_j\}$. Different decision
makers, of course, can classify as optimal different prospects. Since
the utility factor is an objective quantity, we assume that it is the same for 
all decision makers. The difference between the decisions of the pool members
happens because the attraction factors, being subjective quantities, can be
different for different decision makers. Here, we thus do not need to invoke
random utilities and heterogeneous expectations in the objective utility factor
(Cohen, 1980; McFadden and Richter, 1991; Clark, 1995; Regenwetter, 2001).
The heterogeneity or differences between different decision makers appears 
due to the presence of the attraction factor that embodies different states 
of minds among the human population, and as a function of context and time.

The experimental probability that a prospect $\pi_j$ is chosen can be 
defined as a frequency in the following way. Let $M_j$ agents from the 
total number $M$ of decision makers choose the prospect $\pi_j$. Then, 
assuming a large number of agents, the aggregate probability of this 
prospect is given by the frequency 
\be
\label{44}
 p_{exp}(\pi_j) = \frac{M_j}{M} \;  .
\ee
This experimental probability is to be compared with the theoretical 
prospect probability $p(\pi_j)$, using the standard tools of statistical 
hypothesis testing. 

It is also possible to define the aggregate value of the attraction factor 
by the equation
\be
\label{45}
 q(\pi_j) = p_{exp}(\pi_j) - f(\pi_j) \;  
\ee
and to compare this with the mean values $\pm 1/4$.

In this way, QDT provides a practical scheme that can be applied to realistic
problems for various kinds of decision making in psychology, economics, 
finance, and other cases, when behavioral effects are important.  

As an illustration, we have applied this theory to several examples 
in which the disjunction effect occurs. The latter is specified by Savage (1954) 
as a violation of the sure-thing principle. A typical setup for illustrating the
disjunction effect is a two-step gamble (Tversky and Shafir, 1992). Suppose that 
a group of people accepted a gamble in which the player can either win an amount 
of money or lose a possibly different amount. After the first gamble, the 
participants are invited to gamble a second time, being free to either accept the 
second gamble or to refuse it. Experiments by Tversky and Shafir (1992) showed 
that the majority of people accept the second gamble when they know the result 
of the first gamble, whatever its result, whether they won or lost in the 
previous gamble, but only a minority accepted the second gamble when the outcome 
of the first gamble was unknown to them.

Another example, studied by Tversky and Shafir (1992), had to do with a 
group of students who reported their preferences about buying a nonrefundable 
vacation, following a tough university test. They could pass the exam or fail. 
The students had to decide whether they would go on vacation or abstain. It 
turned out that the majority of students purchased the vacation when they 
passed the exam as well as when they had failed. However, only a minority
of participants purchased the vacation when they did not know the results of 
the examination. 

Another example of the disjunction effect concerns stock markets, as 
analyzed by Shafir and Tversky (1992). Consider the USA presidential election, 
when either a Republican or a Democrat wins. On the eve of the election, market
players can either sell certain stocks from their portfolio or hold them. It is
known that a majority of people would be inclined to sell their stocks, if they 
would know who wins, regardless of whether the Republican or Democrat candidate 
wins the upcoming election. This is because people expect the market to fall 
after the elections. At the same time, a great many people do not sell their 
stocks before knowing who really won the election, thus contradicting the 
sure-thing principle. Thus, investors could have sold their stocks before
the election at a higher price, but, abiding to the disjunction effect, they 
were waiting until after the election to know its result, thereby selling 
sub-optimally at a lower price after stocks have already fallen.

We have presented a detailed analysis of the above experiments
(Yukalov and Sornette, 2009b; 2011). The absolute value of the aggregate 
attraction factor (\ref{45}) was found, within the typical statistical error 
of the order of $20\%$ characterizing these experiments, to coincide with the 
predicted value $0.25$.

Another known paradox in classical decision making is the conjunction error.
A typical situation is when people judge about a person, who can possess one 
characteristic and also some other characteristics, as in the often-cited
example of Tversky and Kahneman (1980): "Linda is 31 years old, single, 
outspoken, and very bright. She majored in philosophy. As a student, she was 
deeply concerned with issues of discrimination and social justice, and also 
participated in anti-nuclear demonstrations. Which is more likely? 
(i) Linda is a bank teller; (ii) Linda is a bank teller and is active in the 
feminist movement." Most people answer (ii) which is an example of the 
conjunction fallacy.

There are many other examples of the conjunction fallacy. For a quantitative 
analysis, we have taken the data from Shafir et al. (1990), who present
one of the most carefully accomplished and thoroughly discussed set of 
experiments on the conjunction fallacy. Again, we found 
(Yukalov and Sornette, 2009b; 2011) that the value of the aggregate attraction 
factor, within the experimental accuracy of $20\%$, coincides with $0.25$, in 
excellent agreement with the QDT quarter law. 

The planning paradox has also found a natural explanation within QDT
(Yukalov and Sornette, 2009a). Moreover, it has been shown 
(Yukalov and Sornette, 2010b) that QDT explains practically all typical 
paradoxes of classical decision making, arising when decisions are taken 
by separate individuals.

\section{Influence of information obtained through social interactions}

The standard setup displaying the paradoxes in classical decision making 
corresponds to {\it individual} decision makers that take decisions without 
consulting each other. As has been mentioned in the Introduction, in a number 
of experimental studies, it has been found that exchange of information 
through consultations sharply reduces errors in decision making compared
with the prescription of classical utility theory. For instance, the errors 
in the disjunction effect and conjunction fallacy strongly decrease, when 
group members get information by learning from their experience 
(K\"{u}hberger et al., 2001) or exchange information by consulting 
(Charness et al., 2010).

The theory developed in the previous sections has been formulated for
a decision maker that is a member of a society. An individual decision maker
is just a particular instance for the application of the theory. The suggested 
general approach can also be applied to the case of a decision maker interacting 
with other members of the society and receiving information from them, which 
may change his/her preferences and decrease the errors typical of individual 
decision makers. A decision maker, receiving information from the surrounding 
members of his/her society, can be called a {\it learning decision maker}.

\subsection{Learning decision maker}

Let us denote by $\mu$ a measurable amount of information received by a 
decision maker from the surrounding society. The amount of information can 
be measured by invoking some of the known information measures (Khinchin, 1957;
Arndt, 2004). For instance, information can be represented in the form of a 
Kullback-Leibler information functional over a set of given facts. The 
information can be received through direct interactions, that is, consultations 
with other members of the society. Or each member of the society can receive 
information by learning the results of other agents activity. Thus, the aggregate 
trades of agents in a market produce the data characterizing this market that 
is then available to all and mediates the indirect interactions between them. 
Learning these data gives information to each of the traders (Barber et al., 2009).     

If each member of the society gets the same amount of information $\mu$, 
the state of each member changes, hence the state of the society also varies 
depending on the amount of this additional information. The statistical state, 
characterizing the society, is now a function $\hat{\rho}_{AB}(\mu)$, 
which is normalized as
\be
\label{46}
 {\rm Tr}_{AB} \hat\rho_{AB}(\mu) = 1 \;  .
\ee
We then follow a procedure similar to that described in Section 3. The 
prospect probability is defined as before by
\be
\label{47}
 p(\pi_j,\mu) \equiv  
{\rm Tr}_{AB} \hat\rho_{AB}(\mu)\hat P(\pi_j) \;  ,
\ee
with the difference that we have now an additional variable $\mu$ characterizing
the amount of additional information. By convention, if the latter is set to zero, 
we return to the same formulas as those presented in Section 3. Since the prospect operators act on the space of mind $\mathcal{H}_A$, by defining the reduced 
statistical operator
\be
\label{48}
 \hat\rho_A(\mu) \equiv  {\rm Tr}_{B} \hat\rho_{AB}(\mu) \; ,
\ee
the prospect probability takes the form
\be
\label{49}
 p(\pi_j,\mu) \equiv  
{\rm Tr}_{A} \hat\rho_{A}(\mu)\hat P(\pi_j) \;   .
\ee
And in the matrix representation, we get
\be
\label{50}
 p(\pi_j,\mu) = \sum_{mn} \rho_{mn}(\mu) P_{nm}(\pi_j) \;  ,
\ee
with the notation
\be
\label{51}
 \rho_{mn}(\mu) \equiv \lgl \; m | \; \hat\rho_A(\mu)\; | n \; \rgl  
\ee
and the normalization condition
\be
\label{52}
 {\rm Tr}_{A} \hat\rho_{A}(\mu) = \sum_n \rho_{nn}(\mu) = 1 \; .
\ee

Let us introduce the evolution operator $\hat{U}(\mu)$ that 
describes the evolution of the system state under the varying amount 
of additional information $\mu$. The initial state, before the 
information exchange starts, is
\be
\label{53}
 \hat\rho_{AB}(0) = \hat\rho_{AB} \;  ,
\ee
and corresponds to the situation when decision makers were still separate 
non-interacting individuals. 

The transformation resulting from the interactions between decision makers 
can be represented as
\be
\label{54}
 \hat\rho_{AB}(\mu) = \hat U(\mu) \hat\rho_{AB} \hat U^+(\mu) \;  .
\ee
To satisfy the initial condition (\ref{53}), it is necessary that the
initial value of the evolution operator $ \hat U(0)$
be the identity operator $\hat{1}_{AB}$ acting on $\mathcal{H}_{AB}$:
\be
\label{55}
 \hat U(0) = \hat 1_{AB}   ~.
\ee
In order for the normalization condition (\ref{46}) to be valid for all 
$\mu$, the evolution operator has to be unitary such that
\be
\label{56}
  \hat U^+(\mu)  \hat U(\mu) = \hat 1_{AB} \; .
\ee
Assuming that $\hat{U}(\mu)$ is continuous with respect to $\mu$, 
differentiating condition (\ref{56}), applying the operator $\hat U(\mu)$
and using again (\ref{56}) gives
\be
\label{57}
  \frac{d\hat U(\mu)}{d\mu} + 
\hat U(\mu)\; \frac{d\hat U^+(\mu)}{d\mu} \; \hat U(\mu) = 0 \; .
\ee
This equation for a unitary operator $\hat{U}(\mu)$ can be rewritten as
\be
\label{58}
 i \; \frac{d\hat U(\mu)}{d\mu} = \hat H_{AB} \hat U(\mu) \;  ,
\ee
where $H_{AB}$ is called the {\it evolution generator}, which is a self-adjoint
operator on $\mathcal{H}_{AB}$ assumed to be invariant with respect to $\mu$. 
Equation (\ref{58}) yields the evolution operator 
\be
\label{59}
 \hat U(\mu) = \exp\left ( - i \hat H_{AB} \mu \right ) \;  .
\ee
This evolution operator, in view of Eq. (\ref{54}), defines the variation of 
the total state of the society under the varying amount of information $\mu$.

\subsection{Decision maker as a personality}

The interaction of the decision maker with his/her social environment is 
supposed to ensure that he/she keeps his/her distinct identity and personality 
while, at the same time, possibly changing his/her state of mind. In other 
words, the surrounding society does influence the decision maker state, but 
does so in a way that does not suppress him/her as a person taking his/her 
own decisions. In modeling terms, this corresponds to the behavior of a 
subsystem that is part of a larger system that changes the subsystem 
properties, while the subsystem is not destroyed and retains its typical 
features. Such a subsystem is called {\it quasi-isolated} 
(Yukalov, 2011, 2012a). Another correspondence is the influence 
exerted on a finite system by an external measuring device that acts so as 
not to destroy the main system features, a situation referred to as 
{\it nondestructive measurements} (Yukalov, 2012b). In mathematical language, 
these properties are formulated as follows.  

Reflecting the fact that the total system, that is considered, consists 
of the decision maker, his/her surrounding society, and their mutual 
interactions, the evolution generator $\hat{H}_{AB}$ is represented as a sum 
of the corresponding three terms   
\be
\label{60}
 \hat H_{AB} = \hat H_A + \hat H_B + \hat H_{int} \;  .
\ee
The first term characterizes the decision maker, which implies that the 
operator $\hat{H}_A$ generates the space of mind $\cH_A$ by defining the basis 
of elementary prospects that are typical of the decision maker, through the
eigenvalue problem
\be
\label{61}
 \hat H_A \; | \; n \; \rgl = E_n \; | \; n \; \rgl \;  ,
\ee
with the span over the basis yielding the space of mind (\ref{1}). The second 
term, acting on the space $\cH_B$, describes the surrounding society. And the 
third term, acting on the total space $\cH_{AB}$, corresponds to the 
interaction of the decision maker with his/her social environment, associated 
with the process of information flow. 

As mentioned above, the interaction of the decision maker with his/her social 
environment is supposed to ensure that the decision maker keeps his/her 
identity and personality, although possibly changing his/her state of mind. 
In mathematical language, this is formulated as the following commutativity 
property
\be
\label{62}
 \left [ \hat H_A , \; \hat H_{int} \right ] = 0 \;  ,
\ee
where $[a,b] \equiv ab - ba$. This property, in combination with (\ref{60}), 
is equivalent to the commutativity condition
\be
\label{63}
 \left [ \hat H_A , \; \hat H_{AB} \right ] = 0 \;  .
\ee
Actually, these general properties are sufficient for characterizing the 
decision maker as a distinct personality, and more detailed structure of the 
evolution generators is not important.  

Let the space $\mathcal{H}_B$ be generated by the generator $\hat{H}_B$ 
through the span over the basis formed by the eigenvectors given by the 
eigenproblem
\be
\label{64}
 \hat H_B \; | \; k \; \rgl = B_k \; | \; k \; \rgl \;  .
\ee
In view of Eq. (\ref{62}), there exists a set of real numbers $\{ \bt_{nk}\}$ 
such that the interaction term satisfies the eigenproblem 
\be
\label{65}
 \hat H_{int} \; | \; nk \; \rgl = \bt_{nk} \; | \; nk \; \rgl \;  ,
\ee
in which $|nk\rangle \equiv |n\rangle \otimes |k\rangle$ denotes
the tensorial product between the eigenvectors $|n\rangle$ and $|k\rangle$.
Then, the generator of the total system yields the eigenproblem
\be
\label{66}
 \hat H_{AB} \; | \; nk \; \rgl = 
( E_n + B_k + \bt_{nk} ) \; | \; nk \; \rgl \; .
\ee
  
The above equations make it straightforward to derive the explicit expression
for the prospect probability (\ref{50}). For this purpose, let us introduce some
convenient notations. We define the eigenvalue differences
\be
\label{67}
 \om_{mn} = E_m - E_n \; , \qquad 
\ep_{mnk} \equiv \bt_{mk} - \bt_{nk}  
\ee
and the matrix elements
\be
\label{68}
\overline\rho_{mn}(\mu) = \rho_{mn}(0) \exp ( - i \om_{mn}\mu ) \; ,
\ee
in which
$$
\rho_{mn}(0) = \lgl \; m \; | \; \hat\rho_A \; | \; n \; \rgl =
\rho_{mn} \;   .
$$

We introduce the {\it effect density} describing the distribution of the 
impacts of the surrounding environment affecting the considered decision 
maker during the process characterizing the transfer of information:
\be
\label{69}
 g_{mn}(\ep) \equiv \sum_k \; 
\frac{\lgl\; mk\; |\;\hat\rho_{AB}\;|\; nk\;\rgl}{\lgl\; m\; |\;\hat\rho_A\; |\; n \;\rgl} \; 
\dlt(\ep - \ep_{mnk}) \;  .
\ee
It is clear that the latter is normalized as
\be
\label{70}
  \int_{-\infty}^{+\infty} g_{mn}(\ep) \; d\ep = 1 \; .  
\ee
The Fourier transform of the effect density gives the {\it decoherence factor}
\be
\label{71}
 D_{mn}(\mu) \equiv 
\int_{-\infty}^{+\infty} g_{mn}(\ep) e^{-i\ep\mu}\; d\ep \; .  
\ee

Finally, we come to the prospect probability (\ref{50}) represented as
\be
\label{72}
 p(\pi_j,\mu) = f(\pi_j) + q(\pi_j,\mu) \;  .
\ee
Here, the first term is the same utility factor as in Eq. (\ref{16}). It does 
not depend on the received additional information, being assumed to be 
an objective invariant quantity. And the second term is the attraction factor 
as a function of the received information 
\be
\label{73}
 q(\pi_j,\mu) = \sum_{m\neq n} 
\overline\rho_{mn}(\mu) P_{nm}(\pi_j) D_{mn}(\mu) \;  .
\ee

Generally, the decoherence factor can depend on the indices $m,n$. For 
simplicity, it is possible, resorting to the theorem of average, to employ 
an averaged decoherence factor not depending on the indices, which reduces 
the attraction factor (\ref{73}) to the form
$$
 q(\pi_j,\mu) = \overline q(\pi_j,\mu) D(\mu) \;  ,
$$
where
$$
\overline q(\pi_j,\mu) \equiv 
\sum_{m\neq n} \overline\rho_{mn}(\mu) P_{nm}(\mu) \; .
$$

Since the effect density is normalized as in Eq. (\ref{70}), the 
decoherence factor $D(\mu)$, derived from the above use of the theorem 
of average, with (\ref{71}), enjoys the property $D(0) = 1$. Therefore,
at zero information, the attraction factor 
\be
\label{74}
q(\pi_j,0) = q(\pi_j) 
\ee
has the properties described in Section 3.6, and we return to the initial
prospect probability
\be
\label{75}
 p(\pi_j,0) = p(\pi_j)\;  ,
\ee
defined by Eq. (\ref{18}). 

Thus, we see that the absolute value of the attraction factor essentially 
depends on the value of the decoherence factor (\ref{71}).

\subsection{Attraction factor attenuation}

If the surrounding society does not influence the decision maker, the 
effect density is given by the delta function $\delta(\varepsilon)$. 
Then, the decoherence factor is constant: $D(\mu) = 1$. That is, we 
always have the same expression of the prospect probability as in 
Eq. (\ref{18}), which is quite clear, since getting no additional 
information does not change the preferences of the decision maker. 

The nontrivial situation is when the decision maker consults with other 
members of the society, acquiring additional information. Interactions
of the decision maker with the society can be of different types, which
defines particular forms of the effect density.

If the number of the members in the society is large, and they act on the 
decision maker independently, then, by the central limit theorem, the 
effect density can be modeled by a Gaussian
\be
\label{76}
  g(\ep) =\frac{1}{\sqrt{2\pi}\;\gm} \; \exp \left ( - \;
\frac{\ep^2}{\gm^2} \right ) \; ,
\ee
where $\gamma$ is the variance of the impacts from different members of the 
society. Respectively, the decoherence factor (\ref{71}) is also a Gaussian
\be
\label{77}
 D(\mu) =  exp \left ( - \; \frac{\mu^2}{2\mu^2_c} \right ) 
\qquad \left ( \mu_c \equiv \frac{1}{\gm} \right ) \; ,
\ee
diminishing with the increasing amount of information. Thence, the 
attraction factor (\ref{73}) decreases with increasing $\mu$, which 
implies the decrease of deviations from the classical decision making and 
the attenuation of the related paradoxes, as has been observed in experiments 
with social groups (Charness et al., 2010). The characteristic critical 
decoherence information $\mu_c$ is smaller for larger variance of the 
impacts, when there are many society members with different properties.

When the society members are not independent, the effect density can differ 
from the Gaussian form. For example, it can be given by the Lorentz 
distribution
\be
\label{78}
 g(\ep) =\frac{\gm}{\pi(\ep^2+\gm^2)} \;   .
\ee
As a result, the decoherence factor is exponential:
\be
\label{79}
 D(\mu) = \exp(-\gm \mu) \; .
\ee
  
If the effect density is represented by the Poisson distribution
\be
\label{80}
 g(\ep) = 
\frac{1}{2\gm} \; \exp \left ( - \; \frac{|\ep|}{\gm} \right ) \;  ,
\ee
the decoherence factor is of the power law form:
\be
\label{81}
 D(\mu) = \frac{1}{1+(\gm\mu)^2} \;  .
\ee

When the society influence is described by a uniform distribution
on the bounded interval $[-\gamma, +\gamma]$,
\be
\label{82}
 g(\ep) = \frac{1}{2\gm} \Theta(\gm-\ep) \Theta(\gm+\ep) \;  ,
\ee
where $\Theta(\cdot)$ is a unit-step function, then the decoherence 
factor decays with oscillations as
\be
\label{83}
 D(\mu) = \frac{\sin(\gm\mu)}{\gm\mu} \;  .
\ee

These examples can be generalized by showing that, typically, the decoherence 
factor asymptotically diminishes with increasing information, which leads to
a decreasing attraction factor and a convergence of the prospect probability  
to the classical form characterized by the utility factor. This is summarized 
by the following theorem.

\vskip 2mm

{\bf Proposition 3}. {\it Let the effect density $g(\varepsilon)$ be a 
measurable function. Then, the prospect probability $p(\pi_j,\mu)$, under 
asymptotically large amount of information $\mu$, tends to the classical form 
represented by the utility factor} $f(\pi_j)$: 
\be
\label{84}
 \lim_{\mu\ra\infty} p(\pi_j,\mu) = f(\pi_j) \;  .
\ee

\vskip 2mm

{\it Proof}: Suppose the effect density $g(\varepsilon)$ is measurable, hence 
being not of the delta-function type. By definition (\ref{70}), it is 
$L^1$-integrable. Therefore, by the Riemann-Lebesgue lemma 
(Bochner and Chandrasekharan, 1949), the decoherence factor (\ref{71}) tends 
to zero for asymptotically large $\mu$:    
\be
\label{85}
 \lim_{\mu\ra\infty} D(\mu) = 0\;  .
\ee
Consequently, because of relation (\ref{73}), the attraction factor also
tends to zero:
\be
\label{86}
  \lim_{\mu\ra\infty} q(\pi_j,\mu) = 0 \;  .
\ee
Then, from Eq. (\ref{72}) it follows that the prospect probability reduces
to the classical utility factor, as is stated in Eq. (\ref{84}).

\subsection{Conjunction fallacy disappearance}

Charness et al. (2010) accomplished a series of experiments designed to test 
whether and to what extent individuals succumb to the conjunction fallacy. 
They used an experimental design of Tversky and Kahneman (1983) and found that,
when subjects are allowed to consult with other subjects, the proportions of 
individuals who violate the conjunction principle fall dramatically, 
particularly when the size of the group rises. It has also been found that
financial incentives for providing the correct answer are effective in inducing 
individuals to make efforts to find the correct answer. When individuals are 
forced to think, they recover in their minds additional information that has
been forgotten or shadowed by emotions. The amount of received information
increases with the size of the group. As a result, there is a substantially 
larger drop in the error rate when the group size is increased from two to 
three than when it is increased from one to two (Charness et al., 2007a; 2010).
These findings confirm the earlier studies by Sutter (2005), who finds only 
a marginal difference between the choices of individuals and two-person groups, 
but a significant difference between the choices of two-person and four-person 
groups in an experimental guessing game. In any event, the effects of group 
interaction are not proportional to group size. In other words, the 
error attenuation decays faster than the inverse information, which is 
compatible with the decoherence factors of the Gaussian (\ref{71}) or 
exponential (\ref{73}) forms. In order to determine the exact form of the 
error attenuation, it would be necessary to perform a number of experiments 
in which the group size or the amount of received information would be varied 
over significantly larger intervals than done until now.

In the experiment by Charness et al. (2010), with groups of three members, 
the fraction of individuals giving incorrect answers dropped to $0.17$. It would 
be interesting to study how the errors would diminish with further increase of the 
number of the consulting decision makers. It is clear that the error should not 
disappear completely, since the amount of received information is never actually 
infinite. However, there exists a critical amount of information, when the error 
could be neglected. This critical value would also be interesting to find 
experimentally.

\section{Conclusion}

We have considered the role of information received by decision makers, 
when the agents interact with each other by increasing the amount of their 
mutual information. For this purpose, we have generalized the quantum 
decision theory (QDT), developed earlier for individual decision makers 
(Yukalov and Sornette, 2008, 2009a,b,c, 2010a,b, 2011), to the case of 
decision makers that are members of a society. Mathematically, this corresponds 
to replacing the description of strategic states of decision makers from wave 
functions to statistical operators. In QDT, a choice is made by choosing the 
prospect that corresponds to the largest probability, each prospect probability 
consisting of two terms, a utility factor and an attraction factor. The utility 
factor characterizes the objective utility of prospects, while the attraction 
factor represents subjective feelings, emotions, and biases. Setting the 
attraction factor to zero reduces QDT to the classical decision making based 
on the maximization of expected utility, albeit within a probabilistic framework. 
So, classical decision theory is a 
particular case of QDT. Real decision makers depart from the predictions of classical utility 
theory, which does not take account of the attraction factor, 
leading to a variety of paradoxes. But in QDT, all those paradoxes 
find simple and natural explanations, since the theory accounts for the 
decisions of real human beings.

At an initial stage, when the decision makers of a given society have not had 
yet sufficient time for mutual interactions to increase their information, the 
attraction factor, quantifying the deviations from classical decision theory, 
is crucially important. Its aggregate absolute value is about $0.25$ on a 
maximum scale of $0$ to $1$ for the choice probabilities. It is therefore highly 
significant. The occurrence of the attraction factor is due to the interference 
of prospects in the decision maker brains. Since, in quantum theory, interference 
is necessarily connected with coherence, it is possible to say that the decision 
maker is in a coherent state.

However, the level of this coherence, and the value of the attraction factor,
essentially depend on the amount of information available to a decision maker.
If, in the process of mutual interactions between the members of the society, the 
amount of information of a decision maker increases, then the attraction factor
diminishes. This can be called the decoherence process. Respectively, the 
prospect probabilities tend to their classical values represented by the utility
factors. This rationalizes experimental findings showing that the deviations from 
classical decision making decrease when agents make decisions after receiving 
additional information, for instance, by consulting with each other 
(Charness et al., 2010; Sung and Choi, 2012; Schultze et al., 2012). 

It is possible to imagine a situation where a decision maker receives wrong, that 
is negative, information from the society members, for instance when cheating on 
this particular individual. In that case, the attraction factor could increase, 
hence, the deviations from classical decision making would rise. It would be 
interesting to perform such experiments with decision makers getting wrong or 
misleading information to calibrate better the effect density functions that are 
central to QDT for interacting individuals.

We would like to stress that the central point distinguishing our approach from 
the classical decision theory is the possibility of taking into account subjective
degrees of freedom of decision makers, such as subconscious feelings and behavioral 
biases. Mathematically, this is achieved by employing the techniques of quantum 
theory for defining the probabilities of prospects. This definition results in the
appearance of quantum effects, such as interference, coherence, and decoherence,  
modeling subconscious feelings and biases. Recall that in the interpretation of 
quantum theory there exists the explanation of its statistical properties as due
to hidden variables. The standard techniques of quantum theory avoid directly 
dealing with these variables, nevertheless effectively taking them into account.
The same concerns the application of quantum techniques to decision making. Using
quantum techniques allows us to effectively take into account such hidden variables 
as subconscious feeling and biases, at the same time avoiding their direct 
description. Briefly speaking, the subjective feelings and biases are the hidden 
variables of quantum decision theory, which are treated by the techniques of 
quantum theory. As a result, in quantum decision theory, there appear two 
characteristics, the utility factor and the attraction factor. The first is an
objective characteristic uniquely defined by the utility of the considered prospects.
And the attraction factor describes the subjective attitude of decision makers
to the prospects. Generally, being a subjective characteristic, the attraction 
factor is a random quantity. However, the employed quantum rules allow us to
find the typical value of this factor, as is explained in Sec. 3.6. The knowledge
of this expected typical value makes it possible to make quantitative predictions
for the decision maker choices.        

The mathematical techniques, we have used, may be not customary for scholars in 
social sciences. Therefore, to clearly state the content of this work, 
we list the concrete questions posed in the present paper and the related given 
answers.

\vskip 2mm

\begin{itemize}

\item
{\it Question}: Why is it necessary to develop a novel approach in decision 
making, generalizing the expected utility theory?

\vskip 2mm
{\it Answer}: Classical utility theory does not take into account behavioral 
biases, which leads to numerous paradoxes in decision making.

\item
{\it Question}: Does quantum decision theory remove the paradoxes of classical 
decision making by individual decision makers and, if so, why? 

\vskip 2mm
{\it Answer}: Quantum decision theory removes the paradoxes of classical 
decision theory by individual decision makers by taking into account behavioral 
biases.

\item
{\it Question}: When does quantum decision theory reduce to the classical decision 
theory based on the notion of expected utility?

\vskip 2mm
{\it Answer}: Neglecting behavioral biases reduces quantum decision theory to
the classical decision theory.

\item
{\it Question}: Is it possible to give quantitative predictions in quantum 
decision theory and, if so, do they agree with empirical observations?

\vskip 2mm
{\it Answer}: Quantum decision theory does give quantitative predictions that
are in excellent agreement with empirical observations.

\item
{\it Question}: Why is it necessary to consider decision makers as members of 
a society?  

\vskip 2mm
{\it Answer}: In a society, decision makers interact with each other, 
varying by this their available information.

\item
{\it Question}: How does the amount of available information influence the 
decisions of social agents? 

\vskip 2mm
{\it Answer}: Increasing the amount of information reduces the role of behavioral 
biases.

\item
{\it Question}: Does the reduction of behavioral biases with information 
increase correspond to empirical observations?

\vskip 2mm
{\it Answer}: The reduction of behavioral biases with information increase 
is in agreement with empirical observations, explaining the experimentally 
studied effects of error attenuation, e.g., in disjunction effect and 
conjunction fallacy. 

\end{itemize}

We claim that our approach of developing a ``quantum decision theory'' 
is of a radically different type from the previous works referring to
the use of quantum methods in cognitive sciences, in the sense that we do 
not fine-tune a model or borrow by analogy some quantum mechanical formalism 
to a specific decision making experiment, as often done in this literature. 
In contrast, we have proposed a general formalism, based on the mathematics 
of Hilbert  spaces that applies to any decision making situation. We stress 
it that the use of Hilbert spaces is the only main link with quantum 
mechanics we employ for the underlying mathematical generalization of 
classical probabilities. Actually, except the functional analysis in the 
Hilbert spaces, we do not involve other techniques of quantum theory and 
our approach does not need and does not touch at all those questionable 
problems of quantum theory interpretations.   
 
One should not confuse the mathematical techniques we employ for calculation
purpose and the problems of interpretation concerning physical effects that 
we do not touch and that have nothing to do with the aim of our approach.

In our previous papers, we have shown how our approach explains the 
main paradoxes and fallacies of decision theory, such as Allais paradox, 
the conjunction fallacy, the disjunction effect, and many more, without 
adjustable parameters. We have also predicted new effects that makes the 
theory falsifiable. 

We would like to emphasize that our approach, not merely qualitatively
explains numerous paradoxes in decision theory, but, as we have shown, 
provides quantitative predictions.  

Summarizing, we would like to stress again the basic difference of our 
approach, as compared to all previous works on this topic: (i) we use only 
the mathematical techniques of Hilbert spaces, but do not invoke other physics 
postulates from quantum theory; (ii) we do not need to touch the 
interpretation problems of quantum mechanics that have nothing to do with our 
theory; (iii) our approach is principally different from all previously 
published works, being general and providing quantitative predictions.

\vskip 5mm

{\bf Acknowledgements}

\vskip 3mm

The authors acknowledge financial support from the Swiss National Science
Foundation. Useful discussions with E.P. Yukalova are appreciated.

\vskip 5mm

\newpage

{\Large{\bf References}}

\vskip 5mm

{\parindent=0pt

\vskip 2mm
Abdellaoui, M., Baillon, A., Placido, L., Wakker, P.P. (2011a).
The rich domain of uncertainty: source functions and their experimental 
implementation.
{\it American Economic Review}, 101, 695--723. 

\vskip 2mm
Abdellaoui, M., Driouchi, A., L' Haridon, O. (2011b).
Risk aversion elicitation: reconciling tractability and bias minimization.
{\it Theory and Decision}, 71, 63--80.

\vskip 2mm
Aerts, D., Aerts, S. (1994).
Applications of quantum statistics in psychological studies of decision processes.
{\it Foundation of Science}, 1, 85--97.

\vskip 2mm
Aliev, R.A., Petrycz, W., Huseynov, O.H. (2012).
Decision theory with imperfect probabilities. 
{\it International Journal of Information Technology and Decision Making},
11, 271--306.

\vskip 2mm
Allais, M. (1953). 
Le comportement de l'homme rationnel devant le risque: critique des 
postulats et axiomes de l'ecole Americaine. 
{\it Econometrica}, 21, 503--546.

\vskip 2mm
Al-Najjar, N.I., Weinstein, J. (2009). 
The ambiguity aversion literature: a critical assessment. 
{\it Economics and Philosophy}, 25, 249--284.
 
\vskip 2mm
Ariely, D. (2008). 
{\it Predictably Irrational}. Harper: New York.

\vskip 2mm
Arndt, C. (2004).
{\it Information Measures}. Springer: Berlin.

\vskip 2mm
Arrow, K.J. (1971). 
{\it Essays in the Theory of Risk Bearing}. Markham: Chicago.

\vskip 2mm
Baaquie, B.E. (2004). 
{\it Quantum Finance}. Cambridge University: Cambridge.

\vskip 2mm
Baaquie, B.E. (2009). 
{\it Interest Rates and Coupon Bonds in Quantum Finance}. 
Cambridge University: Cambridge.

\vskip 2mm
Bagarello, F. (2009).
A quantum statistical approach to simplified stock markets. 
{\it Physica A}, 388, 4397--4406.

\vskip 2mm
Barber, B.M., Odean, T., Zhu, N. (2009).
Systematic noise.
{\it Journal of Financial Markets}, 12, 547--569.

\vskip 2mm
Barra, A., Agliari, E. (2012).
A statistical mechanics approach to Granovetter theory.
{\it Physica A}, 391, 3017--3026.  

\vskip 2mm
Barra, A., Contucci, P. (2010).
Toward a quantitative approach to migrants social interactions.
{\it Europhysics Letters}, 89, 68001. 

\vskip 2mm
Bather, J. (2000).
{\it Decision Theory}. Wiley: Chichester.

\vskip 2mm
Berger, J.O. (1985).
{\it Statistical Decision Theory and Bayesian Analysis}. 
Springer: New York.

\vskip 2mm
Bernoulli, D. (1738).
Exposition of a new theory on the measurement of risk.
{\it Proceedings of Imperial Academy of Sciences of St. Petersburg}, 
5, 175--192. Reprinted in (1954), {\it Econometrica}, 22, 23--36.

\vskip 2mm
Blinder, A., Morgan, J. (2005). 
Are two heads better than one? An experimental analysis of group versus 
individual decision-making. 
{\it Journal of Money and Credit Banking}, 37, 789--811.

\vskip 2mm
Bochner, S., Chandrasekharan, K. (1949).
{\it Fourier Transforms}. Princeton University: Princeton.

\vskip 2mm
Bohr, N. (1933).
Light and life. {\it Nature}, 131, 421--423, 457--459.

\vskip 2mm
Bohr, N. (1958).
{\it Atomic Physics and Human Knowledge}. Wiley: New York.

\vskip 2mm
Brock, W.A., Durlauf, S.N. (2001).
Discrete choice with social interactions. 
{\it Review of Economic Studies}, 68, 235--260.  

\vskip 2mm
Buchanan, J.T. (1982).
{\it Discrete and Dynamic Decision Analysis}. Wiley: Chichester.

\vskip 2mm
Busemeyer, J.R., Wang, Z., Townsend, J.T. (2006).
Quantum dynamics of human decision-making. 
{\it Journal of Mathematical Psychology}, 50, 220--241.

\vskip 2mm
Cadogan, G. (2011).
The source of uncertainty in probabilistic preferences over gambles.
http://ssrn.com/abstract=1971954.

\vskip 2mm
Camerer, C.F., Loewenstein, G., Rabin, R. (Eds.) (2003).
{\it Advances in Behavioral Economics}. Princeton University: Princeton.

\vskip 2mm
Charness, G., Rabin, M. (2002). 
Understanding social preferences with simple tests. 
{\it Quarterly Journal of Economics}, 117, 817--869.

\vskip 2mm
Charness, G., Karni, E., Levin, D. (2007a). 
Individual and group decision making under risk: An experimental study 
of Bayesian updating and violations of first-order stochastic dominance. 
{\it Journal of Risk and Uncertainty}, 35, 129--148.

\vskip 2mm
Charness, G., Rigotti, L., Rustichini, A. (2007b). 
Individual behavior and group membership. 
{\it American Economic Review}, 97, 1340--1352.

\vskip 2mm
Charness, G., Karni, E., Levin, D. (2010).
On the conjunction fallacy in probability judgement: new experimental 
evidence regarding Linda.
{\it Games and Economic Behavior}, 68, 551--556.

\vskip 2mm
Chen, Y., Li, S. (2009). 
Group identity and social preferences. 
{\it American Economic Review}, 99, 431--457.

\vskip 2mm
Chew, S. (1983).
A generalization of the quasilinear mean with applications to the measurement 
of income inequality and decision theory resolving the Allais paradox. 
{\it Econometrica}, 51, 1065--1092.

\vskip 2mm
Chew, S., Epstein, L., Segal, U. (1991).
Mixture symmetry and quadratic utility.
{\it Econometrica}, 59, 139--163. 

\vskip 2mm
Cialdini, R.B. (2001).
The science of persuasion. 
{\it Scientific American}, 284, 76--81.

\vskip 2mm
Clark, S.A. (1995).  
The random utility model with an infinite choice space. 
{\it Economic Theory}, 7, 179--189.

\vskip 2mm
Cohen, M.A. (1980). 
Random utility systems - the infinite case. 
{\it Journal of Mathematical Psychology}, 22, 1--23.

\vskip 2mm
Coleman, A.J., Yukalov, V.I. (2000).
{\it Reduced Density Matrices}. Springer: Berlin.

\vskip 2mm
Contucci, P., Gallo, I., Menconi, G. (2008).
Phase transitions in social sciences: two-populations mean-field theory.
{\it International Journal of Modern Physics B}, 22, 2199--2212. 

\vskip 2mm
Cooper, D., Kagel, J. (2005). 
Are two heads better than one? Team versus individual play in signaling games. 
{\it American Economic Review}, 95, 477--509.

\vskip 2mm
Dakic, B., Suvakov, M., Paterek, T., Brukner, C. (2008).
Efficient hidden-variable simulation of measurements in quantum experiments. 
{\it Physical Review Letters}, 101, 190402.

\vskip 2mm
Durlauf, S.N. (1999).
How can statistical mechanics contribute to social science?
{\it Proceedings of National Academy of Sciences of USA}, 96, 10582--10584. 

\vskip 2mm
Edwards, W. (1955).
The prediction of decision among bets. 
{\it Journal of Experimental Psychology}, 50, 200--204.

\vskip 2mm
Edwards, W. (1962).
Subjective probabilities inferred from decisions. 
{\it Psychological Review}, 69, 109--135.
 
\vskip 2mm
Eisert, J., Wilkens, M. (2000).
Quantum games. 
{\it Journal of Modern Optics}, 47, 2543--2556.

\vskip 2mm
Ellsberg, D. (1961).
Risk, ambiguity, and the Savage axioms. 
{\it Quarterly Journal of Economics}, 75, 643--669.

\vskip 2mm
Fehr, E., Rangel, A. (2011).
Neuroeconomic foundations of economic choice: recent advances.
{\it Journal of Economic Perspectives}, 25, 3--30.

\vskip 2mm
French, S., Insua, D.R. (2000).
{\it Statistical Decision Theory}. Arnold: London.

\vskip 2mm
Friedman, M., Savage, L. (1948).
The utility analysis of choices involving risk.
{\it Journal of Political Economy}, 56, 279--304. 

\vskip 2mm
Gollier, G. (2001).
{\it Economics of Risk and Time}. MIT: Cambridge.

\vskip 2mm
Green J., Jullien, B. (1988).
Ordinal independence in nonlinear utility theory.
{\it Journal of Risk and Uncertainty}, 1, 355--387.

\vskip 2mm
Guo, H., Zhang, J., Koehler, G.J. (2008).
A survey of quantum games. 
{\it Decision Supporting Systems}, 46, 318--332.

\vskip 2mm
Hastings, N.A., Mello, J.M. (1978).
{\it Decision Networks}. Wiley: Chichester.

\vskip 2mm
Hey, J. (1984).
The economics of optimism and pessimism: a definition and some applications. 
{\it Kyklos}, 37, 181--205.

\vskip 2mm
Kahneman, D., Tversky, A. (1979).
Prospect theory: an analysis of decision under risk. 
{\it Econometrica}, 47, 263--291.

\vskip 2mm
Karmarkar, U. (1978).
Subjectively weighted utility: a descriptive extension of the expected 
utility model. 
{\it Organizational Behavior and Human Performance}, 21, 61--72.

\vskip 2mm
Karmarkar, U. (1979).
Subjectively weighted utility and the Allais paradox. 
{\it Organizational Behavior and Human Performance}, 24, 67--72.

\vskip 2mm
Keyl, M. (2002).
Fundamentals of quantum information theory. 
{\it Physics Reports}, 369, 431--548.

\vskip 2mm
Khinchin, A.I. (1957).
{\it Mathematical Foundations of Information Theory}.
Dover: New York.

\vskip 2mm
Kitto, K. (2009).
Science and subjectivity. 
{\it Literary Parintantra}, 1, 18--28.

\vskip 2mm
K\"{u}hberger, A., Komunska, D., Perner, J. (2001).
The disjunction effect: does it exist for two-step gambles?
{\it Organizational Behavior and Human Decision Processes}, 85, 250--264.

\vskip 2mm
Kullback, S., Leibler, R.A. (1951). 
On information and sufficiency. 
{\it Annals of Mathematical Statistics}, 22, 79--86. 

\vskip 2mm
Kullback, S. (1959).
{\it Information Theory and Statistics}. Wiley: New York.

\vskip 2mm
Kydland, F.E., Prescott, E.C. (1977).
Rules rather than discretion: the inconsistency of optimal plans. 
{\it Journal of Political Economy}, 85, 473--492.

\vskip 2mm
Lambert-Mogiliansky, A., Zamir, S., Zwirn, H. (2009).
Type indeterminacy: A model of the Kahneman-Tversky man. 
{\it Journal of Mathematical Psychology}, 53, 349--361.

\vskip 2mm
Landsburg, S.E. (2004).
Quantum game theory. 
{\it American Mathematical Society}, 51, 394--399.

\vskip 2mm
Leaw J.N., Cheong, S.A. (2010).
Strategic insights from playing quantum tic-tac-toe. 
{\it Journal of Physics A}, 43, 455304.

\vskip 2mm
Lindgren, B.W. (1971).
{\it Elements of Decision Theory}. Macmillan: New York.

\vskip 2mm
Loewenstein, G., Rick, S., Cohen, J.D. (2008).
Neuroeconomics.
{\it Annual Review of Psychology}, 59, 647--672.

\vskip 2mm
Loomes, G., Sugden, R. (1982).
Regret theory: an alternative theory of rational choice under uncertainty. 
{\it Economical Journal}, 92, 805--824.

\vskip 2mm
Luce, R.D. (1958).
A probabilistic theory of utility.
{\it Econometrica}, 26, 193--224.

\vskip 2mm
Machina, M.J. (2008).
Non-expected utility theory. 
In: {\it New Palgrave Dictionary of Economics}. 
Durlauf, S.N., Blume, L.E. (Eds.), Macmillan: New York.

\vskip 2mm
Malevergne, Y., Sornette, D. (2006).
{\it Extreme Financial Risks}. Springer: Heidelberg.

\vskip 2mm
Markovitz, H. (1952).
The utility of wealth. 
{\it Journal of Political Economy}, 60, 151--158.

\vskip 2mm
Marshall, K.T., Oliver, R.M. (1995).
{\it Decision Making and Forecasting}. McGraw-Hill: New York.

\vskip 2mm
Marschinski, R., Rossi, P., Tavoni, M., Cocco, F. (2007).
Portfolio selection with probabilistic utility.
{\it Annals of Operations Research}, 151, 223--239.

\vskip 2mm
McCaffery, E.J., Baron, J. (2006).
Isolation effects and the neglect of indirect effects of fiscal policies. 
{\it Journal of Behavioral Decision Making}, 19, 289--302.

\vskip 2mm
McFadden, D., Richter, M. (1991). 
Revealed  stochastic preference. 
In: {\it Preferences, Uncertainty and Optimality}. 
Chipman, J.S., McFadden, D., Richter, M.K. (Eds.), 
Westview Press: Boulder.

\vskip 2mm
Neumann, J. von (1955).
{\it Mathematical Foundations of Quantum Mechanics}. 
Princeton University: Princeton.

\vskip 2mm
Neumann, J. von, Morgenstern, O. (1953).
{\it Theory of Games and Economic Behavior}. 
Princeton University: Princeton.

\vskip 2mm
Nielsen, M.A., Chuang, I.L. (2000).
{\it Quantum Computation and Quantum Information}. 
Cambridge University: New York.

\vskip 2mm
Pothos, E.M., Busemeyer, J.R. (2010).
A quantum probability explanation for violations of rational decision theory. 
{\it Proceedings of Royal Society B}, 276, 2171--2178.

\vskip 2mm
Pratt, J.W. (1964).
Risk aversion in the small and in the large. 
{\it Econometrica}, 32, 122--136.

\vskip 2mm
Rabin, M. (2000).
Risk aversion and expected-utility theory: a calibration theorem. 
{\it Econometrica}, 68, 1281--1292.

\vskip 2mm
Raiffa, H., Schlaifer, R. (2000).
{\it Applied Statistical Decision Theory}. Wiley: New York.

\vskip 2mm
Regenwetter, M., Marley, A.A.J. (2001). 
Random relations, random utilities and random functions. 
{\it Journal of Mathematical Psychology}, 45, 864--912.

\vskip 2mm
Rivett, P. (1980).
{\it Model Building for Decision Analysis}. Wiley: Chichester.

\vskip 2mm
Rothschild, M., Stiglitz, J. (1970).
Increasing risk: a definition. 
{\it Journal of Economic Theory}, 2, 225--243.

\vskip 2mm
Rothschild, M., Stiglitz, J. (1971).
Increasing risk: its economic consequences. 
{\it Journal of Economic Theory}, 3, 66--84.

\vskip 2mm
Safra, Z., Segal, U. (2008).
Calibration results for non-expected utility theories. 
{\it Econometrica}, 76, 1143--1166.

\vskip 2mm
Savage, L.J. (1954).
{\it The Foundations of Statistics}. Wiley: New York.

\vskip 2mm
Schultze, T., Mojzisch, A., Schulz-Hardt, S. (2012).
Why groups perform better than individuals at quantitative judgement tasks.
{\it Organizational Behavior and Human Decision Processes}, 118, 24--36.

\vskip 2mm
Seetharaman, P.B. (2003).
Probabilistic versus random-utility models of state.
{\it International Journal of Research in Marketing}, 20, 87--96.

\vskip 2mm
Segal, W., Segal, I.E. (1998).
The Black-Scholes pricing formula in the quantum context. 
{\it Proceedings of National Academy of Sciences of USA}, 95, 4072--4075.

\vskip 2mm
Shafir, E.B., Smith, E.E., Osherson, D.N. (1990).
Typicality and reasoning fallacies. 
{\it Memory and Cognition}, 18, 229--239.

\vskip 2mm
Shafir, E., Tversky, A. (1992). 
Thinking through uncertainty: Nonconsequential reasoning and choice. 
{\it Cognitive Psychology}, 24, 449--474.

\vskip 2mm
Shi, Y. (2009).
Current research trend: information technology and decision making.
{\it International Journal of Information Technology and Decision Making},
8, 1--5.

\vskip 2mm
Simon, H.A. (1955).
A behavioral model of rational choice.
{\it Quarterly Journal of Economics}, 69, 99--118.

\vskip 2mm
Simon, H.A. (1956).
Rational choice and the structure of the environment.
{\it Psychological Review} 63, 129--138. 
 
\vskip 2mm
Sornette, D. (2003).
{\it Why Stock Markets Crash}. Princeton University: Princeton.

\vskip 2mm
Strotz, R.H. (1955).
Myopia and inconsistency in dynamic utility maximization. 
{\it Review of Economic Studies}, 23, 165--180.

\vskip 2mm
Sung, S.Y., Choi, J.N. (2012).
Effects of team management on creativity and financial performance
of organizational teams.
{\it Organizational Behavior and Human Decision Processes}, 118, 4--13.

\vskip 2mm
Sutter, M. (2005). 
Are four heads better than two? An experimental beauty-contest game with 
teams of different size. 
{\it Economic Letters}, 88, 41--46.

\vskip 2mm
Tapia Garcia, J.M., Del Moral, M.J., Martinez, M.A., Herrera-Viedma, E. (2012).
A consensus model for group decision-making problems with interval fuzzy 
preference relations.
{\it International Journal of Information Technology and Decision Making},
11, 709--725. 

\vskip 2mm
Tversky, A., Kahneman, D. (1980). 
Judgements of and by representativeness. 
In: {\it Judgements Under Uncertainty: Heuristics and Biases}. 
Kahneman, D., Slovic, P., Tversky, A. (Eds.), 
Cambridge University: New York, p. 84--98.

\vskip 2mm
Tversky, A., Kahneman, D. (1983).
Extensional versus intuitive reasoning: the conjunction fallacy in 
probability judgement. 
{\it Psychological Review}, 90, 293--315.

\vskip 2mm
Tversky, A., Shafir, E. (1992).
The disjunction effect in choice under uncertainty. 
{\it Psychological Science}, 3, 305--309.

\vskip 2mm
Weirich, P. (2001).
{\it Decision Space}. Cambridge University: Cambridge.

\vskip 2mm
West,  B.J., Grigolini, P. (2010).
A psychophysical model of decision making. 
{\it Physica A}, 389, 3580--3587.

\vskip 2mm
White, D.I. (1976).
{\it Fundamentals of Decision Theory}. Elsevier: New York.

\vskip 2mm
Williams, C.P., Clearwater, S.H. (1998).
{\it Explorations in Quantum Computing}. Springer: New York.

\vskip 2mm
Xu, Z. (2011).
Approaches to multi-stage multi-attribute group decision making.
{\it International Journal of Information Technology and Decision Making},
10, 121--146.

\vskip 2mm
Yaari, M. (1987).
The dual theory of choice under risk. 
{\it Econometrica}, 55, 95--115. 

\vskip 2mm
Yukalov, V.I. (2002).
Stochastic instability of quasi-isolated systems.
{\it Physical Review E}, 65, 056118.

\vskip 2mm
Yukalov, V.I. (2003a).
Irreversibility of time for quasi-isolated systems.
{\it Physics Letters A}, 308, 313--318.

\vskip 2mm
Yukalov, V.I. (2003b).
Expansion exponents for nonequilibrium systems.
{\it Physica A}, 320, 149--168.

\vskip 2mm
Yukalov, V.I. (2007).
Representative ensembles in statistical mechanics.
{\it International Journal of Modern Physics}, 21, 69--86.

\vskip 2mm
Yukalov, V.I. (2011).
Equilibration and thermalization in finite quantum systems.
{\it Laser Physics Letters}, 8, 485--507.

\vskip 2mm
Yukalov, V.I. (2012a).
Equilibration of quasi-isolated quantum systems.
{\it Physics Letters A}, 376, 550--554.

\vskip 2mm
Yukalov, V.I. (2012b).
Decoherence and equilibration under nondestructive measurements.
{\it Annals of Physics (N.Y.)}, 327, 253--263.

\vskip 2mm
Yukalov, V.I., Sornette, D. (2008).
Quantum decision theory as quantum theory of measurement. 
{\it Physics Letters A}, 372, 6867--6871.

\vskip 2mm
Yukalov, V.I., Sornette, D. (2009a).
Physics of risk and uncertainty in quantum decision making. 
{\it European Physical Journal B}, 71, 533--548.

\vskip 2mm
Yukalov, V.I., Sornette, D. (2009b).
Processing information in quantum decision theory. 
{\it Entropy}, 11, 1073--1120.

\vskip 2mm
Yukalov, V.I., Sornette, D. (2009c).
Scheme of thinking quantum systems. 
{\it Laser Physics Letters}, 6, 833--839.

\vskip 2mm
Yukalov, V.I., Sornette, D. (2010a).
Entanglement production in quantum decision making. 
{\it Physics of Atomic Nuclei}, 73, 559--562.

\vskip 2mm
Yukalov, V.I., Sornette, D. (2010b).
Mathematical structure of quantum decision theory. 
{\it Advances in Complex Systems}, 13, 659--698.

\vskip 2mm
Yukalov, V.I., Sornette, D. (2011).
Decision theory with prospect interference and entanglement. 
{\it Theory and Decision}, 70, 283--328.

\vskip 2mm
Yukalov,V.I. Sornette, D. (2013).
Quantum probabilities of composite events in quantum measurements with
multimode states.
{\it Laser Physics}, 23, 105502. 
 
\vskip 2mm
Zabaleta, O.G., Arizmendi, C.M. (2010).
Quantum dating market.
{\it Physica A}, 389, 2858--2863.

\end{document}